\documentclass[10pt,table,a4paper]{article} 

\usepackage[hidelinks]{hyperref}

\usepackage{amsmath}
\usepackage{amssymb}

\usepackage{siunitx}

\usepackage{xcolor}

\definecolor{azure}{rgb}{0.0, 0.5, 1.0}
\definecolor{asparagus}{rgb}{0.53, 0.66, 0.42}
\definecolor{amaranth}{rgb}{0.9, 0.17, 0.31}
\definecolor{cadetgrey}{rgb}{0.57, 0.64, 0.69}

\usepackage{tikz}
\usetikzlibrary{shapes.geometric, arrows,matrix}
\usetikzlibrary{calc}
\usetikzlibrary{fit}
\tikzstyle{startstop} = [rectangle, rounded corners, minimum width=3cm, minimum height=0.5cm,text centered, draw=black, fill=cadetgrey!30]
\tikzstyle{io} = [trapezium, trapezium left angle=70, trapezium right angle=110, minimum width=3cm, minimum height=0.5cm, text centered, draw=black] 
\tikzstyle{process} = [rectangle, minimum width=3cm, minimum height=0.5cm, text centered, draw=black] 
\tikzstyle{decision} = [diamond, minimum width=2cm, minimum height=0.25cm, text centered, draw=black] 
\tikzstyle{arrow} = [thick,->,>=stealth]

\usepackage{listings}

\usepackage{caption}

\usepackage{algorithm}
\usepackage{algorithmicx}
\usepackage{algpseudocode}

\usepackage{nomencl}
\makenomenclature

\usepackage{todonotes}

\usepackage{xfrac}

\algdef{SE}{HPXLoop}{EndHPXLoop}[1]{\textcolor{azure}{\textbf{parallel\_for}} \(\mbox{#1}\) \textcolor{azure}{\textbf{do}}}{\textcolor{azure}{\textbf{end}}}
\algnewcommand\algorithmicforeach{\textbf{for each}}
\algdef{S}[FOR]{ForEach}[1]{\algorithmicforeach\ #1\ \algorithmicdo}

\numberwithin{equation}{section}

\newcommand{\sref}[2]{\hyperref[#2]{#1 \ref*{#2}}}
\newcommand{\R}{\mathbb{R}}  

\newcommand{\yellow}[1]{#1}
\newcommand{\future}[1]{\textbf{\textcolor{amaranth}{future}}~#1 =}
\newcommand{\sync}[1]{\textbf{\textcolor{amaranth}{wait}}\{#1\} }

\usepackage{subcaption}
\usepackage{booktabs}
\usepackage{a4wide}

\usepackage[numbers,sort&compress]{natbib}

\begin{document}

\title{An asynchronous and task-based implementation of Peridynamics utilizing HPX -- the C++ standard library for parallelism and concurrency}

\author{Patrick Diehl$^{1,3,4,*}$, Prashant K. Jha$^5$, Hartmut Kaiser$^{3,4}$, Robert Lipton$^{2}$, \\ and Martin Levesque$^1$ \\
1. Department for Multi scale Mechanics, Polytechnique Montreal \\
2. Department of Mathematics, Louisiana State University\\
3. Center of Computation \& Technology, Louisiana State University\\
4. Ste$\vert\vert$ar group \\
5. Oden Institute for Computational Engineering and Sciences,\\ The University of Texas at Austin \\
* patrickdiehl@lsu.edu (https://orcid.org/0000-0003-3922-8419)\\
}

\maketitle
\thispagestyle{empty}

\begin{abstract}
On modern supercomputers, asynchronous many task systems are emerging to address the new architecture of computational nodes. Through this shift of increasing cores per node, a new programming model with focus on handling of the fine-grain parallelism with increasing amount of cores per computational node is needed. Asynchronous Many Task (AMT) run time systems represent a paradigm for addressing the fine-grain parallelism. They handle the increasing amount of threads per node and concurrency. HPX, a open source C++ standard library for parallelism and concurrency, is one AMT which is conforming to the C++ standard.
Motivated by the impressive performance of asynchronous task-based parallelism through HPX to N-body problem and astrophysics simulation, in this work, we consider its application to the Peridynamics theory. Peridynamics is a non-local generalization of continuum mechanics tailored to address discontinuous displacement fields arising in fracture mechanics. Peridynamics requires considerable computing resources, owing to its non-local nature of formulation, offering a scope for improved computing performance via asynchronous task-based parallelism. 
Our results show that HPX based Peridynamic computation is scalable and the scalability is in agreement with the theory. 
In addition to the scalability, we also show the validation results and the mesh convergence results. For the validation, we consider implicit time integration and compare the result with the classical continuum mechanics (CCM) (Peridynamics under small deformation should give similar results as CCM). For the mesh convergence, we consider explicit time integration and show that the results are in agreement with theoretical claims in previous works.
\end{abstract}



\section{Introduction}
Modern supercomputers' many core architectures, like \yellow{field-programmable gate arrays (FPGAs)} and Intel Knights Landing, provide more threads per computational node as before~\cite{sutter2005free,ross2008cpu}. Through this shift of increasing cores per node, a new programming model with the focus on handling of the fine-grain parallelism with increasing amount of cores per computational node is needed.\\

Asynchronous Many Task (AMT)~\cite{thoman2018taxonomy} run time systems represent an emerging paradigm for addressing the fine-grain parallelism since they handle the increasing amount of threads per node and concurrency~\cite{4777912}. Existing task-based programming models can be classified into three classes: (1) Library solutions, \emph{e.g.}\ StarPU~\cite{starpu}, Intel TBB~\cite{inteltbb}, Argobots~\cite{Seo2018ArgobotsAL}, Qthreads~\cite{qthreads}, Kokkos~\cite{CarterEdwards20143202}, and HPX~\cite{Kaiser2020} (2) language extensions, \emph{e.g.}\ Intel Cilk Plus~\cite{intelcilkplus} and OpenMP $4$.$0$~\cite{openmpspec}, and (3) programming languages, \emph{e.g.}\ Chapel~ \cite{chamberlain07parallelprogrammability}, Intel ISPC~\cite{intelispc}, and X$10$~\cite{Charles:2005:XOA:1094811.1094852}. \\

Most of the tasked-based libraries are based on C or C++ programming language. The C++ $11$ programming language standard  laid the foundation for concurrency by introducing futures, that have shown to support the concept of futurization to enable a task-based parallelism~\cite{cxx11_standard}. In addition, the support for parallel execution with the so-called parallel algorithms were introduced in the C++ $17$ standard~\cite{cxx17_standard}.\\

HPX is an open source asynchronous many task run time system that focuses on high performance computing~\cite{Kaiser2020}. HPX provides wait-free asynchronous execution and futurization for synchronization. It also features parallel execution policies utilizing a task scheduler, which enables a fine-grained load balancing parallelization and synchronization due to work stealing. HPX's application programming interface (API) is in strict adherence to the C++ $11$~\cite{cxx11_standard} and C++ $17$ standard API definitions~\cite{cxx17_standard}. For example for the concept of futurization the \lstinline|hpx:future| can be replaced by \lstinline|std::future| without breaking the API.\\

The HPX library was recently utilized to parallelize a N-Body code using the asynchronous task-based implementation and was compared against non-AMT implementations~\cite{khatami2016massively,heller2013using}. In many cases the HPX implementation outperformed the MPI/OpenMP implementation. 
The HPX-library has been utilized in an astrophysics application describing the time evolution of a rotating star. Test run on a Xeon Phi resulted in a speedup by factor of two with respect to a $24$-core Skylake SP platform~\cite{Pfander:2018:AOS:3204919.3204938}. On the NERSC's Cori super-computer the simulation of the Merger of two stars could achieve $96.8$\% parallel efficiency on $648$,$280$ Intel Knight's landing cores~\cite{operationbell17} and $68.1$\% parallel efficiency on 2048 cores~\cite{dai2019piz} of Swiss National Supercomputing Centre's Piz Daint super-computer. In addition, a speedup up to $2.8$x was shown on full system run on Piz Daint by replacing the Message Passing Interface (MPI) with libfabric~\cite{grun2015brief} for communication. Motivated by the acceleration seen for these problems, we consider the application of HPX to peridynamics theory.

Peridynamics is a non-local generalization of continuum mechanics, tailored to address discontinuous displacement fields that arise in fracture mechanics~\cite{Diehl2019review,javili2019peridynamics}. 
Several peridynamics implementations utilizing the EMU nodal discretization~\cite{CMPer-Parks} are available. Peridigm~\cite{Parks2012} and PDLammps~\cite{CMPer-Parks} rely on the widely used MPI for the inter-node parallelization. Other approaches rely on acceleration cards, like OpenCL~\cite{MOSSAIBY20171856} and CUDA~\cite{Diehl:2012}, for parallelization. These single device GPU approaches, however, cannot deal with large node clouds due to current hardware memory limitations on GPUs.\\ 

In peridynamics, each point in domain interacts with neighboring points within specified non-local length scale $\delta$ referred to as horizon. In numerical discretization, the mesh size is typically much smaller than the horizon, meaning that the mesh nodes, in addition to depending on the adjacent nodes of common finite element, depends on the mesh nodes within distance $\delta$ (horizon). This non-local nature of calculation makes the peridynamics based simulation slow as compared to the local formulation such as linear elastodynamics. 
To reduce the computational burden, several versions of local to non-local coupling methods, in which the non-local calculations are restricted to smaller portions of domain and elsewhere local calculations arising from elastodynamics formulation, have been proposed~\cite{d2019review}. 
While these methods reduce the cost considerably, the problem of efficient compute resource utilization for non-local calculations remains. \\

This work presents two peridynamics models discretized with the EMU nodal discretization making use of the features of AMT within HPX. We show in \yellow{this work} how to take advantage of the fine-grain parallelism arising on modern supercomputers. In addition, The speed-up $\mathsf{S}$ and the parallel efficiency $\mathsf{E}$ are shown in Figure~\ref{fig:benchmark:im2} and Figure~\ref{fig:benchmark:ex2}. The implementation is validated against results from classical continuum mechanics and numerical analysis to show that the novel task-based implementation is correct.\\

The paper is structured as follows. Section~\ref{sec:hpx} briefly introduces HPX and associated key concepts utilized in the asynchronous task-based implementation. Section~\ref{sec:background} reviews peridynamics models, and the EMU-ND discretization. Section~\ref{sec:coding} describes the proposed modular design and the implementation using HPX. Section~\ref{sec:validation} presents the validation and the mesh convergence results. 
The actual computational time for the HPX peridynamics implementation is benchmarked against the theoretical computation time in Section~\ref{sec:benchmark}. Section~\ref{sec:conclusion} concludes this work.

\section{HPX -- an open source C++ standard library for parallelism and concurrency}
\label{sec:hpx}
\begin{figure}[tb]
\centering
\begin{tikzpicture}
\draw [draw=black] (0,0) rectangle (5,4);
\node[align=center,font=\large] at (2.5,4.25) {HPX};
\node[fit={(0.5,0.25) (4.5,1.25)}, inner sep=0pt, draw=black] (parcel) {};
\node[align=center,font=\large] at (parcel.center) {Parcel Handler};
\node[fit={(0.5,1.5) (4.5,2.5)}, inner sep=0pt, draw=black] (agas) {};
\node[align=center,font=\large] at (agas.center) {AGAS};
\node[fit={(0.5,2.75) (4.5,3.75)}, inner sep=0pt, draw=black] (thread) {};
\node[align=center,font=\large] at (thread.center) {Thread Manager};
\end{tikzpicture}
\caption{Run time components of HPX: Thread manager, Active Global Address Space (AGAS), and the Parcel handler. The thread manager provides a high-level API for the task-based programming and deals with the light-weight user level threads. The Active Global Address Space (AGAS) provides global identifiers for all allocated memory. The parcel handler provides the communication between different computational nodes in the cluster.}
\label{fig:hpx:components}
\end{figure}
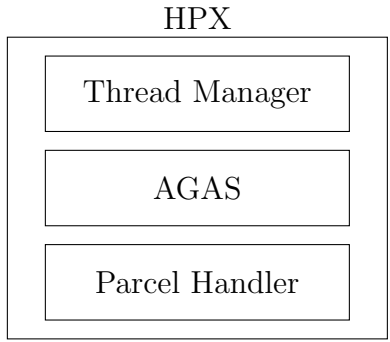

The HPX library~\cite{Kaiser2020} is a C++ standard compliant Asynchronous Many Task (AMT) run time system tailored for high performance computing (HPC) applications. Figure~\ref{fig:hpx:components} shows the three components provided by HPX. First, the \textit{thread manager}~\cite{kaiser2014hpx}, which provides a high-level API for the task-based programming and deals with the light-weight user level threads. Second, the \textit{Active Global Address Space (AGAS)}~\cite{kaiser2014hpx}, which provides the global identifiers to hide the explicit message passing. So the access of a remote or local object is unified. Third, the \textit{parcel handler}~\cite{ac:2017,kaiser2009parallex} for communication between computational nodes in the cluster, which provides calls to remote computational nodes in a C++ fashion. The communication between the nodes is realized either by the Transmission Control Protocol (tcp) protocol or the message passing Interface (MPI). For more implementation details about these components, we refer to~\cite{Kaiser2020} and for a general overview we refer to~\cite{thoman2018taxonomy}.

HPX provides well-known concepts such as data flow, futurization, and Continuation Passing Style (CPS), as well as new and unique overarching concepts. The combination of these concepts results in a unique model. The concept of futurization and parallel for loops, which are used to synchronize and parallelize, are recalled here.  

\subsection{Futurization}
An important concept for synchronization provided by HPX is futurization. The API provides an asynchronous return type \lstinline$hpx::lcos::future<T>$. This return type, a so-called future, is based on templates and hides the function call return type. The difference here is that the called function immediately returns, even if the return value is not computed. The return value of a future is accessible through the \lstinline$.get()$ operator that is an synchronous call and waits until the return type is computed. The standard-conforming API functions \lstinline$hpx::wait_all$, \lstinline$hpx::wait_any$, and \lstinline$hpx::lcos::future<T>::then$ are available for combining futures and generate parallel executions graphs~\cite{Kaiser:2015:HPL:2832241.2832244}.\\

\begin{table}[tbp]
\begin{tabular}{cc}
\raisebox{.1cm}{\begin{minipage}{0.325\textwidth}
\begin{tikzpicture}
\node (a) [draw,circle] at (0,2) {$H$};
\node (b) [draw,circle] at (0,0) {$P$};
\node (c) [draw,circle] at (1,0) {$X$};
\draw [->] (a) -- (b);
\draw [->] (a) -- (c);
\end{tikzpicture}

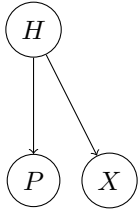
\captionof{figure}{Example dependency graph.}
\label{fig::hpx::depengraph}
\end{minipage}} &
\begin{minipage}{0.625\textwidth}
\begin{lstlisting}[language=c++,caption=Modeling the dependency graph in Figure~\ref{fig::hpx::depengraph} with composition in HPX.,label=lst::hpx::example]
//Vector with all dependencies of h
std::vector<hpx::lcos::future<void>> dependencies;
dependcy.push_back(compute(p));
dependcy.push_back(compute(x));
hpx::wait_all(dependencies);
compute(h,p,x);
\end{lstlisting}
\end{minipage}
\end{tabular}
\end{table}

A typical example dependency graph is shown in Figure~\ref{fig::hpx::depengraph}. On the Figure, $H$ depends asynchronously on $P$ and $X$. Listing~\ref{lst::hpx::example} provides these dependencies resolutions within HPX. First, a \lstinline$std::vector$ is utilized to collect the dependencies. Second, the computations futures are added to this vector. Note, that the \lstinline$compute$ function returns a future in both cases, immediately, and the computations are executed asynchronously. Third, a barrier with \lstinline$hpx::wait_all$ for the dependencies has to be defined before $H$ can be computed. HPX internally ensures that the function in line $6$ is only called when the previous two futures computations are finished.

\subsection{Parallelization}
Consider the addition of $n$ elements for the two vectors \lstinline$p$ and \lstinline$x$, where the sum is stored piece-wise in vector \lstinline$h$. Listing~\ref{lst::cpp::example} shows the sequential approach for this computation while Listing~\ref{lst::hpx::example2} shows the same computational task but the sum is executed in parallel. The \lstinline$for$ loop is replaced with \lstinline$hpx::parallel::for_loop$ which is conforming with the C++ $17$ standard~\cite{cxx17_standard}. In other words, \lstinline$hpx::parallel::for_loop$ can be replaced by \lstinline$std::for_each$ which is currently an experimental feature in the GNU compiler collection $9$ and Microsoft VS $2017$ $15.5$. The first argument defines the execution policy while the second and third define the loop range. The last argument is the lambda function, executed in parallel for all $i$ ranging from $0$ to $z$. Note that only two lines of codes are required to execute the code in parallel. The parallel for loop can be combined with futurization for synchronization. Therefore, the parallel execution policy is changed to \lstinline$hpx::parallel::execution::par(hpx::parallel::execution::task)$.\\

\begin{lstlisting}[language=c++,caption=C++ code for storing the sum of two vectors sequentially in third vector.,label=lst::cpp::example,float,floatplacement=ptb]
//Sequential loop 
for (size_t i=0,i<z;i++)
{
   h[i] = p[i]+x[i];	
} 
\end{lstlisting}
\begin{lstlisting}[language=c++,caption=HPX equivalent code for storing the sum of two vectors parallel in a third vector.,label=lst::hpx::example2,float,floatplacement=ptb]
//Synchronizing parallel for loop
hpx::parallel::for_loop(
hpx::parallel::execution::par,
0,z,[h,p,x](boost::uint64_t i)
{
   h[i] = p[i]+x[i];
});
\end{lstlisting}

The future can now be synchronized with other futures. Listing~\ref{lst::hpx::example3} shows an example for synchronization. Vectors \lstinline$p$ and \lstinline$x$ are independently manipulated before the pairwise sum is computed. Therefore, the execution policy is changed and the futures of the manipulations are synchronized with the third future in line $26$. Here, the \lstinline$hpx::wait_all$~ensures that the manipulations are finished and \lstinline$then$ describes the sum's dependencies.

\begin{lstlisting}[language=c++,caption=Example for the synchronization of three parallel for loops within HPX by using the concept of futurization.,label=lst::hpx::example3,float,floatplacement=p]
std::vector<hpx::lcos::future<void>> dep;

dep.push_back(hpx::parallel::for_loop(
hpx::parallel::execution::par(
hpx::parallel::execution::task),
0,z,[p](boost::uint64_t i)
{
   p[i] = p[i]+1;
});
dep.push_back(hpx::parallel::for_loop(
hpx::parallel::execution::par(
hpx::parallel::execution::task),
0,z,[x](boost::uint64_t i)
{
   x[i] = x[i]-1;
});)

hpx::lcos::future f = hpx::parallel::for_loop(
hpx::parallel::execution::par(
hpx::parallel::execution::task),
0,z,[h,p,x](boost::uint64_t i)
{
   h[i] = p[i]+x[i];
});

hpx::wait_all(dep).then(f);

\end{lstlisting}

\section{Peridynamics theory}
\label{sec:background}
In this section, we briefly introduce the key feature of peridynamics theory essential for the implementation. For more details about PD we refer to~\cite{javili2019peridynamics} and for the utilized material models to~\cite{CMPer-Silling,CMPer-Lipton3,CMPer-Lipton}.

\begin{figure}[tb]
\centering
\begin{tikzpicture}
\draw (0,0) circle (1);
 \node at (0,2.25) {{\small $\Omega_0$}};
 \draw[fill=black!100](0,0) circle (2pt);
 \node at (0,-0.25) {{\small $\mathbf{X}$}};
 \node at (-0.5,-0.25) {{\small $\mathbf{X}'$}};
 \draw (-0.5,0.) -- (0,0);
 \draw (0,0) ellipse (3cm and 2cm);
\draw [->] (0,0) to (0.95,0.225);
\node at (1.125,0.25) {{\tiny $\delta$}};
\node at (1.1,0.8) {{\tiny $B_\delta(\mathbf{X})$}};
\draw[fill=gray!80] (0,0.5) circle (2pt);
\draw[fill=gray!80] (0,-0.5) circle (2pt);
\draw[fill=gray!80] (0.5,0.0) circle (2pt);
\draw[fill=gray!80] (-0.5,0.0) circle (2pt);
\node at (3.5,3) {{\small Mechanical load}};
\draw [arrow, bend angle=45, bend left,->]  (2,2) to (5,2);
\draw (6,0) ellipse (2.5cm and 1.5cm);
\node at (6.125,1.75) {{\small $\Omega_t$}};
\draw[fill=gray!80] (5.55,0.0) circle (2pt);
\draw[fill=black!100](6,0) circle (2pt);
 \node at (6.5,-0.25) {{\small $\mathbf{x}(t,\mathbf{X})$}};
 \node at (4.75,0) {{\small $\mathbf{x}(t,\mathbf{X}')$}};
 \draw (6,0) -- (5.55,0.0);
\end{tikzpicture}
\caption{Left: Schematic representation of PD where the material point $\mathbf{X}\in\Omega$ interacts non-locally with all other material points inside $B_\delta(\mathbf{X})$. Right: Positions $\mathbf{x}(t,\mathbf{X})$ and $\mathbf{x}(t,\mathbf{X}')$ of the material points in the configuration $\Omega_t$.}
\label{fig:principle:pd}
\end{figure}
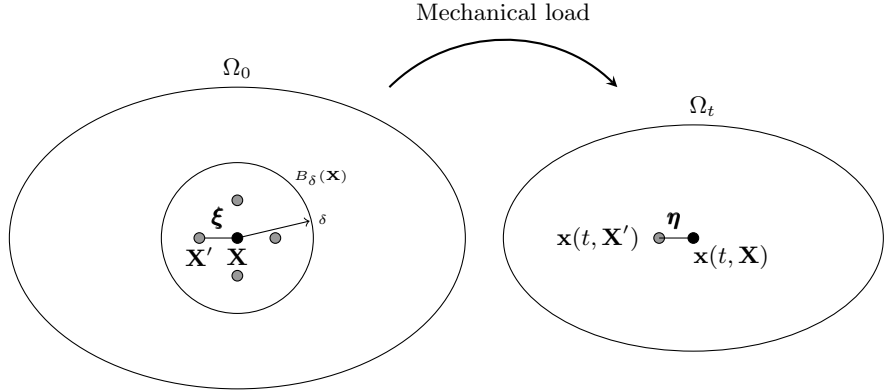

Let a material domain be $\Omega_0\subset \R^d$, for $d= 1,2,$ and $3$. Peridynamics (PD)~\cite{silling2007peridynamic,CMPer-Silling} assumes that every material point $\mathbf{X}\in \Omega_0$ interacts non-locally with all other material points inside a horizon of length $\delta > 0$, as illustrated in Figure~\ref{fig:principle:pd}. Let $B_{\delta}(\mathbf{X})$ be the sphere of radius $\delta$ centered at $\mathbf{X}$. 
When $\Omega_0$ is subjected to mechanical loads, the material point $\mathbf{X}$ assumes position $\mathbf{x}(t,\mathbf{X}) = \mathbf{X} + \mathbf{u}(t,\mathbf{X})$, where $\mathbf{u}(t,\mathbf{X})$ is the displacement of material point $\mathbf{X}$ at time $t$. \\ 

Let $\mathbf{f}(t,\mathbf{u}(t,\mathbf{X}')-\mathbf{u}(t,\mathbf{X}), \mathbf{X}'-\mathbf{X})$ denote the peridynamic force as a function of time $t$, bond-deformation vector $\mathbf{u}(t,\mathbf{X})-\mathbf{u}(t,\mathbf{X})$, and reference bond vector $\mathbf{X}-\mathbf{X}'$. The peridynamics equation of motion is given by
\begin{align}
\varrho(\mathbf{X}) \ddot{\mathbf{u}}(t,\mathbf{X}) = \int\limits_{B_\delta(\mathbf{X})} \mathbf{f}(t,\mathbf{u}(t,\mathbf{X}')-\mathbf{u}(t,\mathbf{X}), \mathbf{X}'-\mathbf{X})d\mathbf{X}' + \mathbf{b}(t,\mathbf{X}),  \label{eq:motion:bond}
\end{align}
where $\mathbf{b}$ is the external force density, \yellow{and} $\varrho(\mathbf{X})$ is the material's mass density. Equation~\eqref{eq:motion:bond} is complemented by initial conditions $\mathbf{u}(0,\mathbf{X}) = \mathbf{u}_0(\mathbf{X})$ and $\dot{\mathbf{u}}(0,\mathbf{X}) = \mathbf{v}_0(\mathbf{X})$. In contrast to local problems, boundary conditions in peridynamics are defined over layer or collar $\Omega_c$ surrounding the domain $\Omega_0$. Boundary conditions will be described in later sections when describing the numerical experiments. 

The  model in \eqref{eq:motion:bond} depends on \yellow{two-point} interactions and is referred to as a bond-based peridynamics model. Bond based model can only model the material with Poisson's ratio $0.25$~\cite{Kunin2011,Kunin2012}. On the other hand, state-based peridynamics models~\cite{silling2007peridynamic} allow for multi-point non-local interactions and overcomes the restriction on the Poisson's ratio.  It is conveniently formulated in terms of displacement dependent tensor valued functions. Let $\underline{T}[t,\mathbf{X}]$ be the peridynamic state at time $t$ and material point $\mathbf{X}$. The peridynamics equation of motion for a state-based model is given by 
\begin{align}
\varrho(\mathbf{X}) &\ddot{\mathbf{u}}(t,\mathbf{X}) = \nonumber\\ 
&\int\limits_{B_\delta(\mathbf{X})} \left(\underline{T}[\mathbf{X},t]\langle \mathbf{X}'-\mathbf{X} \rangle - \underline{T}[\mathbf{X}',t]\langle \mathbf{X} - \mathbf{X}'\rangle \right) d\mathbf{X}' + \mathbf{b}(t,\mathbf{X})
\text{.}\label{eq:motion:state}
\end{align}
Equation~\eqref{eq:motion:state} is complemented by initial conditions $\mathbf{u}(0,\mathbf{X}) = \mathbf{u}_0(\mathbf{X})$ and $\dot{\mathbf{u}}(0,\mathbf{X}) = \mathbf{v}_0(\mathbf{X})$. 

\subsection{Discretization of peridynamics equations}
Continuous and discontinuous Galerkin finite element methods~\cite{chen2011continuous}, Gauss quadrature~\cite{weckner2005numerical} and spatial discretization~\cite{emmrich2007peridynamic,CMPer-Parks} are different discretization approaches for PD. 
Owing to its efficient load distribution scheme, the method of Silling and Askari~\cite{silling2005meshfree,CMPer-Parks}, the so-called EMU nodal discretization (EMU ND) is chosen for this implementation.  \\ 

\begin{figure}[tp]
\centering
\begin{tikzpicture}[scale=1.5]
 \foreach \x in {1,2,3}
 \foreach \y in {1,2,3}
 \shade[ball color = black](\x,\y) circle (2pt);
 \draw[darkgray](1.5,1.5) -- (2.5,1.5) -- (2.5,2.5) -- (1.5,2.5) -- (1.5,1.5);
 \node at (2.25,1.75) { $\mathbf{x}_5$};
 \draw[<-,darkgray] (2.25,2.25) -- (2.65,2.5);
 \draw(1,1) -- (1,3);
 \draw(2,1) -- (2,3);
 \draw(3,1) -- (3,3);
 \node[darkgray] at (2.75,2.5) { $V_5$};
 \draw(1,1) -- (3,1);
 \draw(1,2) -- (3,2);
 \draw(1,3) -- (3,3);
\end{tikzpicture}
\caption{The initial positions of the discrete nodes  $\mathbf{X}:=\lbrace \mathbf{X}_i \in \mathbb{R}^2 : i=1,\ldots,9\rbrace$ in the reference configuration $\Omega_0$. For the discrete node $\mathbf{x}_5$ its surrounding volume $V_5$ is shown schematically.}
\label{fig:dual:mesh}
\end{figure}
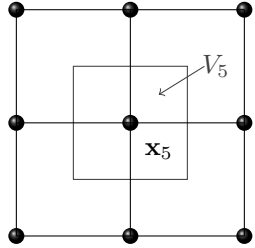

In the EMU ND scheme, the reference configuration is discretized and the discrete set of material points $\mathbf{X}:=\lbrace \mathbf{X}_i \in \mathbb{R}^d : i=1,\ldots,n\rbrace$ are considered to represent the material domain, see Figure~\ref{fig:dual:mesh}. 
The discrete neighborhood $B_\delta(\mathbf{X}_i)$ of the node $\mathbf{X}_i$ yields $B_\delta(\mathbf{X}_i):=\lbrace \mathbf{X}_j \vert \, \vert \mathbf{X}_j - \mathbf{X}_i \vert \leq \delta \rbrace$. 
Each node $\mathbf{X}_i$ represents $V_i$ volume such that $\sum_{i}^n V_i = V_{\varOmega_0}$, where $V_{\varOmega_0}$ is volume of whole domain. We denote nodal volume vector as $\mathbf{V}$. \\

The discrete bond-based equation of motion  yields, for all $\mathbf{X}_i \in \Omega_0$,
\begin{align}
\varrho(\mathbf{X}_i) \ddot{\mathbf{u}}(t,\mathbf{X}_i) = \sum\limits_{B_\delta(\mathbf{X}_i)} \mathbf{f}(t,\mathbf{u}(t,\mathbf{X}_j)-u(t,\mathbf{X}_i),\mathbf{X}_j-\mathbf{X}_i)V_j + \mathbf{b}(t,\mathbf{X}_i) \label{eq:motion:bond:descrite}
\end{align}
and the discrete state-based equation of motion yields, for all $\mathbf{X}_i \in \Omega_0$,
\begin{align}
\varrho(\mathbf{X}_i) \ddot{\mathbf{u}}(t,\mathbf{X}_i) = \sum\limits_{B_\delta(\mathbf{X}_i)} \left(\underline{T}[\mathbf{X}_i,t]\langle \mathbf{X}_j-\mathbf{X}_i \rangle - \underline{T}[\mathbf{X}_j,t]\langle \mathbf{X}_i - \mathbf{X}_j\rangle \right) V_j + \mathbf{b}(t,\mathbf{X}_i)
\text{.}\label{eq:motion:state:descrite}
\end{align}

\textit{Remark: }There could be nodes $\mathbf{X}_j$ in $B_\delta(\mathbf{X}_i)$ such that $V_j$ is partially inside the ball $B_\delta(\mathbf{X}_i)$. For these nodes, we compute the correction in volume, $V_{ij}$, along the lines suggested in [Section 2, \cite{seleson2016convergence}]. In both Equation~\eqref{eq:motion:bond:descrite} and \eqref{eq:motion:state:descrite} we replace $V_j$ by $V_j V_{ij}$ to correctly account for the volume.

\section{Implementation of PeridynamicHPX}
\label{sec:coding}
\noindent
We introduce a modern C++ library, referred to as PeridynamicHPX, that utilizes HPX for parallelism. 
The design is modular and template based making it easy to extend the code with new material models. We present the design of the code and describe the use of parallelism and concurrency based on HPX.

\subsection{Design with respect to modular expandability}
\label{sec:design}
\begin{figure}[ptb]
\includegraphics[width=\linewidth]{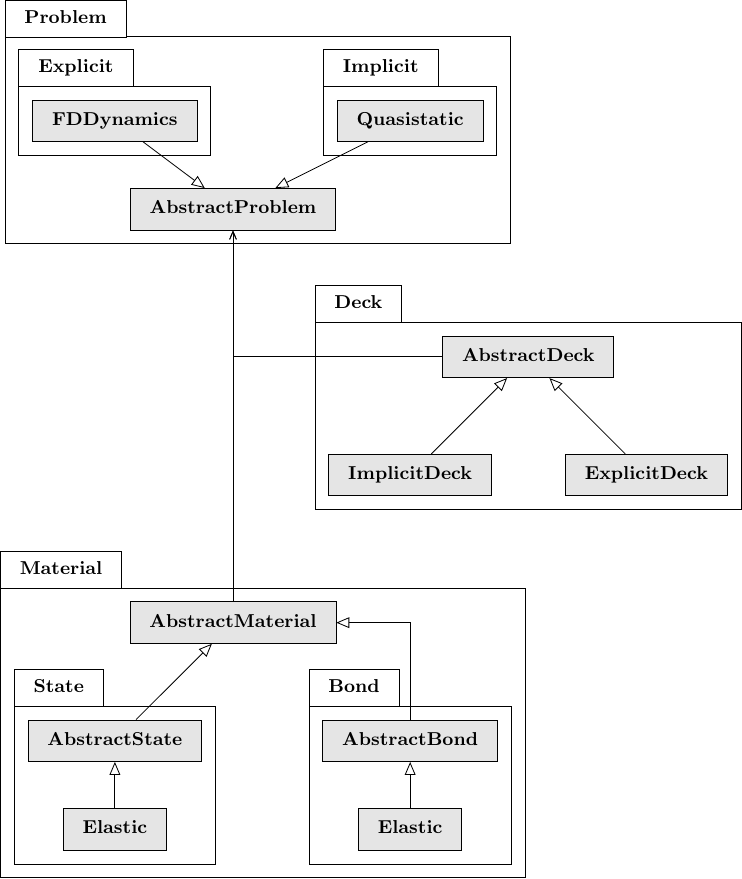}
\caption{Class diagram of PeridynamicHPX which is designed to easily extend the code with new materials or discretizations. The functionality of the code is defined in three packages: \textit{Problem} containing the different classes for discretizations, \textit{Deck} which handles the input and output, and \textit{Material} providing the different kind of material models. All packages provide an abstract class which needs to be inherited by all sub classes for extending the code.}
\label{fig:classdiagram}
\end{figure}

Figure~\ref{fig:classdiagram} shows the modular design class. PeridynamicHPX contains three modules that are affected by the discretization extensions and material models. First, the \textit{Deck} module handles the loading of the discretization and the configuration in the YAML\footnote{http://yaml.org/} file format. Each new Deck \lstinline|class newDeck : public  AbtstractDeck| inherits the common functions from the \textit{AbstractDeck} and is extended with the new problem/material specific attributes.\\

Second, the abstractions for bond-based \lstinline|class AbstractBond| and state-based \lstinline|class AbstractState| materials are provided in the \textit{Material} module. The nonlinear bond-based elastic material \lstinline|class Elastic : public  AbstractBond |  in Section~\ref{sec:material:bond} and the linear state-based elastic material \lstinline|class Elastic : public AbstractState | of Section~\ref{sec:material:state} were implemented. The abstract material must be inherited and the methods, e.g.\ force and strain, are implemented if a new material model is to be implemented. Note that we tried to use as less as possible virtual functions to avoid some overhead which virtual function calls have.\\ 

The different time integration schemes and the discretizations are considered third. All new \textit{Problem} classes inherit their common and abstract functions from \textit{AbstractProblem}. The class \lstinline|template<class T> class  Quasistatic : public   AbstractProblem| implements the implicit time integration in Section~\ref{sec:time:implicit} and class \lstinline|template<class T>class  Explicit : public    AbstractProblem| implements the explicit time integration in Section~\ref{sec:time:explicit}. Note that a problem is a template class with the material definition as the template class. Thus, the specific implementation can therefore be used for state-based and bond-based materials.\\

The design aims to hide most of the HPX specific features and making the code easy for adding new material models by implementing the abstract classes. Listing~\ref{lst::main::example} shows how to run the implicit time integration and the explicit time integration using a elastic state-based material models. Note that we hide the inclusion of header files and parsing the command line arguments. For new problem classes, the user have to deal with the parallel for loops instead of using the C++ standard for loop. Thus, the code is \yellow{accessible to users} that do not have advanced programming experience in parallel computing, but still yields acceptable scalability without optimization.

\begin{lstlisting}[language=c++,caption=Small example of the interfaces and functions to run the implicit time integration and the explicit time integration using a elastic state-based material model. Note that we hide the inclusion of header files and parsing the command line arguments.,label=lst::main::example,float,floatplacement=ptb]
// Include the relevant headers

int main(int argc, char *argv[]) {

// Parse the command line arguments: filename

// Read the YAML config file and store all entities in the deck
IO::deck::ImplicitDeck* deck = new IO::deck::ImplicitDeck(filename);

// Run the implicit time integration with the elastic state material
problem::Quasistatic<material::state::Elastic>* problem =
					new problem::Quasistatic<material::state::Elastic>(deck);
					
// Read the YAML config file and store all entities in the deck
IO::deck::ExplicitDeck* deck2 = new IO::deck::ExplicitDeck(filename);
					
// Run the explicit time integration with the elastic state material
problem::Explicit<material::state::Elastic>* problem =
					new problem::Explicit<material::state::Elastic>(deck2);


return EXIT_SUCCESS;
}
\end{lstlisting}

\subsection{Parallelization with HPX}
\label{sec:parallel}
\noindent
The implementation of bond-based material and the explicit time integration is adapted from earlier one-dimensional code developed by the second author~\cite{CMPer-JhaLipton2}.
The implementation of the state-based material and the implicit time integration is adapted from~\cite{Littlewood2015}. 
These sequential algorithms are analyzed to make use of HPX parallelism. The use of HPX tools to achieve parallelism is discussed in the sequel. 

\subsubsection{Bond-based material}
\label{sec:material:bond}
\noindent
Algorithm~\ref{alg:bondmaterial} shows the use of HPX for computing the internal force and strain energy. For demonstration, we consider nonlinear bond-based peridynamic (NP) model introduced in~\cite{CMPer-Lipton3,CMPer-Lipton}. The same algorithm will work with commonly used bond-based PMB (prototype micro-elastic brittle) model. In Algorithm~\ref{alg:bondmaterial}, $S$ is the bond-strain, $\psi$ is the pairwise potential function, and $J^\delta(r)$ is the influence function. In Section~\ref{sec:validation:ex:2} we show specific form of $\psi$ and $J^\delta$. We read input file and initialize the material class with specified form of function $\psi$ and $J^\delta$. We compute and store the list of neighbors for each node in the reference configuration (initial configuration). HPX parallel for loop is used to compute the force and energy of each node in parallel, see Algorithm~\ref{alg:bondmaterial}.

\begin{algorithm}[tbp]
\caption{Computation of internal force for bond-based material. Here, $S$ is the bond-strain, $\psi$ is the nonlinear potential and $J^\delta(|\mathbf{X}|)$ is the influence function considered in~\cite{CMPer-Lipton3,CMPer-Lipton}.}\label{alg:bondmaterial}

\begin{algorithmic}[1]
\HPXLoop{$i<n$}\Comment{Compute force at mesh nodes}
	\For{$j\in B_\delta(\mathbf{X}_i)$}
        	\State $\pmb{\xi}=\mathbf{X}_j-\mathbf{X}_i$ 
		\State $S=\frac{\mathbf{u}(\mathbf{X}_j)-\mathbf{u}(\mathbf{X}_i)}{|\pmb{\xi}|} \cdot \frac{\pmb{\xi}}{|\pmb{\xi|}}$
		\State\Comment{Compute force at $\mathbf{X}_i$ as in~\cite{CMPer-Lipton3,CMPer-Lipton} }
		\State $\mathbf{f}(\mathbf{X}_i) += \frac{4}{\delta |B_\delta(\mathbf{0})|}J^\delta(|\pmb{\xi}|) \psi'(|\pmb{\xi}|S^2)  S \frac{\pmb{\xi}}{|\pmb{\xi}|} V_j$ \label{alg:bondmaterial:f}
		\State \Comment{Compute strain energy at $\mathbf{X}_i$ as in~\cite{CMPer-Lipton3,CMPer-Lipton}}
		\State $U(\mathbf{X}_i) += \frac{1}{|B_\delta(\mathbf{0})|} J^\delta(|\pmb{\xi}|)\psi(|\pmb{\xi}|S^2)V_j/\delta$ \label{alg:bondmaterial:e}
	\EndFor	
\EndHPXLoop
\end{algorithmic}

\end{algorithm}

\subsubsection{Linearly elastic state-based material}
\label{sec:material:state}
\noindent
The internal force density and strain energy are computed for a state based elastic peridynamic solid material, as described in~\cite{silling2007peridynamic}. Algorithm~\ref{alg:statematerial} shows the adapted algorithm~\cite{Littlewood2015} parallelized and synchronized with HPX. First, the weighted volume $m$ is computed for each node using a HPX parallel for loop. Second, the dilation $\theta$ is computed for each node a HPX parallel for loop. Note that the dilation $\theta$ depends on the weighted volume $m$ and therefore we can not use asynchronous code execution here. The internal force density and the strain energy can be computed independently of each other. Therefore, the execution policy \lstinline$hpx::parallel::execution::task$ is utilized to obtain futures \lstinline|f1| and \lstinline|f2| back of these two loops to compute the loops asynchronously, see Line~\ref{alg:statematerial:f1} and Line~\ref{alg:statematerial:f2}. Note that the strain energy is optional and is only computed if requested, \emph{e.g.}\ for output. A synchronization for these two futures is needed before the force and strain energy are computed in future steps, see Line~\ref{alg:statematerial:wait}.

\begin{algorithm}[tb]
\caption{Computation of internal force and strain energy. Adapted from~\cite{Littlewood2015}}\label{alg:statematerial}
\begin{algorithmic}[1]
\HPXLoop{$i<n$}\Comment{Compute weighted volumes as in~\cite{silling2007peridynamic,Littlewood2015}}
	\State $m_i=0$
	\For{$j\in B_\delta(\mathbf{X}_i)$}
        \State $\pmb{\xi}=\mathbf{X}_j-\mathbf{X}_i$
	\State $m_i += \vert \pmb{\xi} \vert^2 V_j$
      \EndFor	
\EndHPXLoop
\HPXLoop{$i<n$}\Comment{Compute dilatation as in~\cite{silling2007peridynamic,Littlewood2015}}
	\State $\theta_i=0$
	\For{$j\in B_\delta(\mathbf{X}_i)$}
        	\State $\pmb{\xi}=\mathbf{X}_j-\mathbf{X}_i$ 
		\State $\pmb{\eta}=\mathbf{u}(\mathbf{X}_j)-\mathbf{u}(\mathbf{X}_i)$
		\State $\underline{e} = \vert \pmb{\xi} + \pmb{\eta} \vert - \vert \pmb{\xi} \vert$
		\State $\theta_i += \sfrac{3}{m_i}\vert\pmb{\xi}\vert \underline{e}V_j$
      \EndFor	
\EndHPXLoop
\State \future{f1} \label{alg:statematerial:f1}
\HPXLoop{$i<n$}\Comment{Compute internal forces as in~\cite{silling2007peridynamic,Littlewood2015}}
	\For{$j\in B_\delta(\mathbf{X}_i)$}
        	\State $\pmb{\xi}=\mathbf{X}_j-\mathbf{X}_i$ 
		\State $\pmb{\eta}=\mathbf{u}(\mathbf{X}_j)-\mathbf{u}(\mathbf{X}_i)$
		\State $\underline{e}^d = e - \sfrac{(\theta_i\vert \pmb{\xi}\vert)}{3}$
		\State $\underline{t}=\sfrac{3}{m_i}K\theta_i\vert\pmb{\xi}\vert + \sfrac{(15\mu)}{m_i}\underline{e}^d$
	\State $\underline{M} = \sfrac{\pmb{\eta}+\pmb{\xi}}{\vert\pmb{\xi}+\pmb{\eta}\vert}$ 
	\State $\mathbf{f}_i += \underline{t}\underline{M}V_j$
	\State $\mathbf{f}_j -= \underline{t}\underline{M}V_i$
	\EndFor	
\EndHPXLoop
\State \Comment{Note that the strain energy is optional and only computed if requested, e.g. for output}
\State \future{f2} \label{alg:statematerial:f2}
\HPXLoop{{$i<n$}}\Comment{Compute strain energy as in~\cite{silling2007peridynamic}}
	\For{$j\in B_\delta(\mathbf{X}_i)$}
	\State $\pmb{\xi}=\mathbf{X}_j-\mathbf{X}_i$ 
	\State $\pmb{\eta}=\mathbf{u}(\mathbf{X}_j)-\mathbf{u}(\mathbf{X}_i)$
	\State $\underline{M}_{ji} = \sfrac{\pmb{\eta}+\pmb{\xi}}{\vert\pmb{\xi}+\pmb{\eta}\vert}$ 
	\State $\pmb{\xi}=\mathbf{X}_i-\mathbf{X}_j$ 
	\State $\pmb{\eta}=\mathbf{u}(\mathbf{X}_i)-\mathbf{u}(\mathbf{X}_j)$
	\State $\underline{M}_{ij} = \sfrac{\pmb{\eta}+\pmb{\xi}}{\vert\pmb{\xi}+\pmb{\eta}\vert}$ 
	\State $e_i= \sfrac{1}{2} M_{ij} M_{ji} \alpha  V_j $
	\EndFor
\EndHPXLoop
\State \sync{f1,f2} \label{alg:statematerial:wait}
\end{algorithmic}
\end{algorithm}

\subsubsection{Implicit time integration}
\label{sec:time:implicit}
\begin{figure}[ptb]
    \centering
	\begin{tikzpicture}[node distance=1.125cm, scale=0.75, transform shape]
	\node (start) [startstop] {Start}; 
	\node (dec1) [decision, below of=start , yshift=-1.cm] { $s\leq$ steps};
	\node (stop) [startstop, right of=dec1, xshift=3.cm] {Finished}; 
	\node (n0) [process, below of=dec1,yshift=-1.cm] {Update  external force $\mathbf{b}$};
	\node (n1) [process, below of=n0] {Compute residual $\mathbf{r}$~\eqref{eq:residual}};
	\node (dec2) [decision, below of=n1 , yshift=-1.cm] { $\vert \mathbf{r} \vert\geq \tau$ };
	\node (n2) [process, below of=dec2,yshift=-1.cm] { Guess displacement $\mathbf{u}_t$};
	\node (n3) [process, below of=n2,yshift=-0.5cm] { Assembly tangent stiffness matrix $\mathbf{K}$ with $\mathbf{u}_t$};
	\node (n4) [process, below of=n3] { Solve $\mathbf{K}\underline{\mathbf{u}}=-r$};
	\node (n5) [process, below of=n4] { $\mathbf{u}_t = \mathbf{u}_t +\underline{\mathbf{u}}$};
	\node (n6) [process, below of=n5] { Compute residual $\mathbf{r}$};
	\node (dec3) [decision, below of=n6 , yshift=-.5cm] { $\vert \mathbf{r} \vert\geq \tau$ };
	\node (n7) [process, below of=dec3, yshift=-1.0cm] { $\mathbf{u}^{s+1}=\mathbf{u}_t$};
	\node (n8) [process, below of=n7] {$s=s+1$};
	\draw [->] (start) -- (dec1);
	\draw [->] (n0) -- (n1);
	\draw [->] (n1) -- (dec2);
	\draw [->] (n2) -- (n3);
	\draw [->] (n3) -- (n4);
	\draw [->] (n4) -- (n5);
	\draw [->] (n5) -- (n6);
	\draw [->] (n6) -- (dec3);
	\draw [->] (n7) -- (n8);
	\draw [->] (n8.west) -- ++(-3.6,0) -- ++(0,19.) -- ++(5.1,0);
	\draw [->] (dec1) -- node[anchor=north] {no} (stop);
	\draw [->] (dec1) -- node[anchor=west] {yes} (n0);
	\draw [->] (dec2) -| node[anchor=north] {no} (stop);
	\draw [->] (dec2) -- node[anchor=west] {yes} (n2);
	\draw [->] (dec3) -- node[anchor=west] {no} (n7);
	\draw [->] (dec3.east) -- ++(3.8,0) -- ++(0,3) -- ++(0,2) --                
     	node[xshift=1.cm,yshift=-2.5cm, text width=2.5cm]
     	{yes}(n3.east);
	\draw  ($(n3.north west)+(-0.4,0.15)$)  rectangle ($(dec3.south east)+(4.5,-0.6)$);
	\node[above] at ($(n3.north east)+(0.5,0.15)$) {Newton step};
	\end{tikzpicture}
	\caption{Flow chart of the implicit time integration scheme adapted from~\cite{Littlewood2015}. For each time step $s$ the external force $\mathbf{b}$ is updated and the residual $\mathbf{r}$ is evaluated. When the norm of the residual is larger than the tolerance $\tau$ the displacement $\mathbf{u}^{s+1}$ is obtained via a Newton step.}
	\label{fig:algorithm:state:chart}
\end{figure}
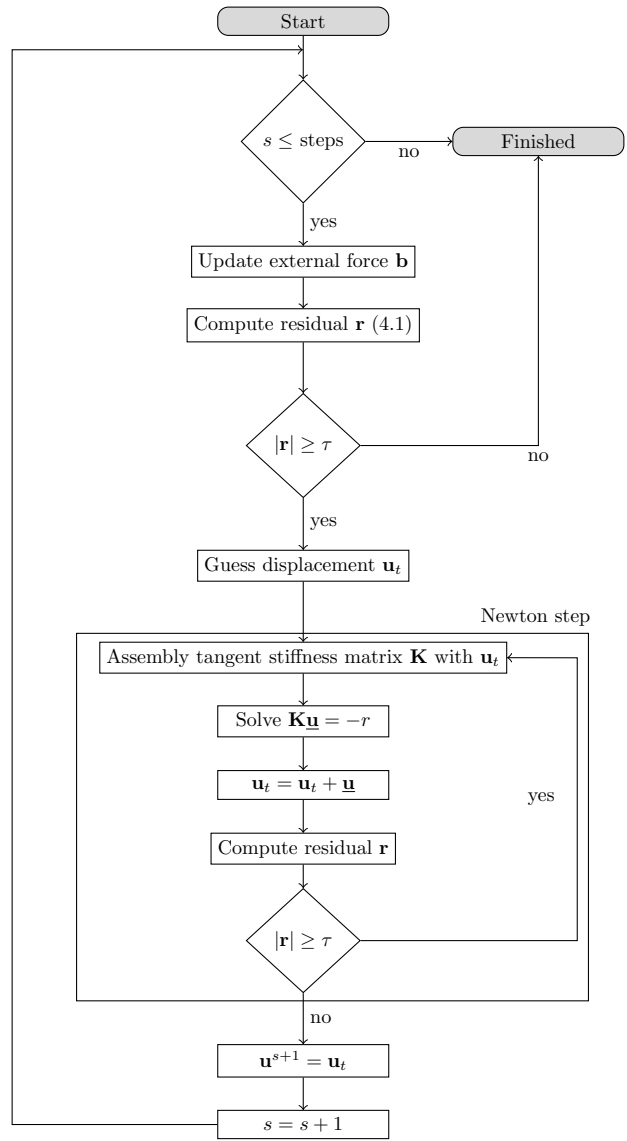

Figure~\ref{fig:algorithm:state:chart} shows the implicit integration implementation flow chart. The external force $\mathbf{b}$ is updated for each load step $s$. Next, the residual
\begin{align}
\mathbf{r} = \sum\limits_{\Omega_0} \mathbf{f}(t,x_i) + \mathbf{b}(t,x_i)
\label{eq:residual}
\end{align}
and its $l_2$ norm are computed and compared against the tolerance $\tau$. If the residual is too large, the tangent stiffness matrix
\begin{align}
\mathbf{K}_{ij} \approx \frac{\mathbf{f}(x_i,u_i+\epsilon^j)-\mathbf{f}(x_i,u_i-\epsilon^j)}{2\epsilon} \textbf{.}
\label{eq:tangent}
\end{align}
is assembled as described in~\cite{Littlewood2015}, (see Algorithm~\ref{alg:tangent:stiffness}). The displacement of the previous load step was used to assembly the first matrix $\mathbf{K}(u)$. Line~\ref{alg:stiffness:force} perturbs the displacement by $\pm\upsilon$, where $\upsilon$ is infinitesimally small. Line~\ref{alg:stiffness:diff} computes the internal forces $f^{\pm\upsilon}$ and Line~\ref{alg:stiffness:matrix} evaluates the central difference to construct the stiffness matrix $\mathbf{K}(u)$. Note that the node's neighborhood $B_\delta$ is represented and has several zero entries where nodes do not have neighbors. Next, the guessed displacement is updated with the solution from solving  $\mathbf{K}\underline{\mathbf{u}} = -\mathbf{r}$. The residual is evaluated \yellow{once} and the Newton method is iterated until $\vert r \vert \leq \tau$. Note that a dynamic relaxation method (DR)~\cite{kilic_peridynamic_nodate}, a conjugate gradient method (CG), or a Galerkin method~\cite{WANG20127730} could be used as well.\\

The high-performance open-source C++ library Blaze~\cite{iglberger2012expression,6266939} was used for matrix and vector operations. Blaze supports HPX for parallel execution and can be easily integrated. The library BlazeIterative\footnote{https://github.com/tjolsen/BlazeIterative} was used for solving $\mathbf{K} \underline{\mathbf{u}}=-\mathbf{r}$. The Biconjugate gradient stabilized method (BiCGSTAB) or the conjugate gradient (CG) solver were used for solving.

\begin{algorithm}[tbp]
\caption{Assembly of the tangent stiffness matrix by central finite difference. Adapted from~\cite{Littlewood2015}}
\label{alg:tangent:stiffness}

\begin{algorithmic}[1]
\State $\mathbf{K}^{d\cdot n\times d\cdot n} = 0$\Comment{Set matrix to zero}

\HPXLoop{$i<n$}
	\For{$i\in \lbrace B_\delta(\mathbf{X}_i),i\rbrace$}
		\State \Comment{Evaluate force state under perturbations of displacement}
		\ForEach{displacement degree of freedom $r$ at node $j$}
			\State $\underline{T}[\mathbf{X}_i](\mathbf{u}+\mathbf{\upsilon}^r)$\label{alg:stiffness:force}
			\State $\underline{T}[\mathbf{X}_i](\mathbf{u}-\mathbf{\upsilon}^r)$
			\For{$k\in B_\delta(\mathbf{X}_i)$}
				\State $\mathbf{f}^{\upsilon +} = \underline{T}^{\upsilon +}\langle \mathbf{X}_k-\mathbf{X}_i\rangle V_i V_j$\label{alg:stiffness:diff}
				\State $\mathbf{f}^{\upsilon -} = \underline{T}^{\upsilon -}\langle \mathbf{X}_k-\mathbf{X}_i\rangle V_i V_j$
				\State $\mathbf{f}_\text{diff} = \mathbf{f}^{\upsilon +} - \mathbf{f}^{\upsilon -}$
				\ForEach{degree of freedom $s$ at node $k$ }
					\State $\mathbf{K}_{sr} += \sfrac{f_{\text{diff}_s}}{2\upsilon}$\label{alg:stiffness:matrix}	
				\EndFor
			\EndFor
		\EndFor
      \EndFor	
\EndHPXLoop

\end{algorithmic}

\end{algorithm}

\subsubsection{Explicit time integration}
\label{sec:time:explicit}
\begin{figure}[ptb]
    \centering
	\begin{tikzpicture}[node distance=1.125cm, scale=0.75, transform shape]
	\node (start) [startstop] {Start}; 
	\node (dec1) [decision, below of=start , yshift=-1.cm] { $k\leq$ steps};
	\node (stop) [startstop, right of=dec1, xshift=3.cm] {Finished}; 
	\node (n0) [process, below of=dec1,yshift=-1.cm] {Update boundary conditions};
	\node (n1) [process, below of=n0] {Compute peridynamics force and external forces $\mathbf{f}^k + \mathbf{b}^k$};
	\node (n2) [process, below of=n1, yshift=-1.cm, xshift=-4cm] {$\mathbf{u}^{k+1} = 2\mathbf{u}^{k} - \mathbf{u}^{k-1} + \Delta t^2 (\mathbf{f}^k + \mathbf{b}^k)$ \eqref{eq:finite:difference}};
	\node (n3) [process, below of=n1, yshift=-1.cm, xshift=4cm] {$\mathbf{v}^{k+1/2} = \mathbf{v}^{k} + (\Delta t/2) (\mathbf{f}^k + \mathbf{b}^k)$ \eqref{eq:velocity:verlet:1}};
	\node (n4) [process, below of=n3, yshift=-1.cm] {$\mathbf{u}^{k+1} = \mathbf{u}^{k} + \Delta t \mathbf{v}^{k+1/2}$ \eqref{eq:velocity:verlet:2}};
	\node (n5) [process, below of=n4,yshift=-1.cm] {Update boundary conditions};
	\node (n6) [process, below of=n5, yshift=-1.cm] {Compute peridynamics force and external forces $\mathbf{f}^{k+1}+\mathbf{b}^{k+1}$};
	\node (n7) [process, below of=n6, yshift=-1.cm] {$\mathbf{v}^{k+1} = \mathbf{v}^{k+1/2} + (\Delta t/2) (\mathbf{f}^{k+1} + \mathbf{b}^{k+1})$ \eqref{eq:velocity:verlet}};
	\node (n8) [io, below of=n7, yshift=-1.cm, xshift=-4.3cm] {Output displacement and velocity at time step $k$};
	\node (n9) [process, below of=n8, yshift=-1.cm] {$k = k+1$};
	\draw [->] (start) -- (dec1);
	\draw [->] (n0) -- (n1);
	\draw [->] (n1) -- (n2);
	\draw [->] (n2) -- ++(0,-9.9) -- (n8);
	\draw [->] (n1) -- (n3);
	\draw [->] (n3) -- (n4);
	\draw [->] (n4) -- (n5);
	\draw [->] (n5) -- (n6);
	\draw [->] (n6) -- (n7);
	\draw [->] (n7) -- (n8);
	\draw [->] (n8) -- (n9);
	\draw [->] (n9.west) -- ++(-6.0,0) -- ++(0,20.2) -- ++(6.3,0);	
	\draw [->] (dec1) -- node[anchor=north] {no} (stop);
	\draw [->] (dec1) -- node[anchor=west] {yes} (n0);	
	\draw [->] (n1) -- node[anchor=east] {central difference scheme} (n2);
	\draw [->] (n1) -- node[anchor=west] {velocity verlet scheme} (n3);	
	\end{tikzpicture}
	\caption{Flow chart of the explicit time integration scheme. For each time step $k$ the boundary conditions are updated and the internal and external forces are computed. Depending on a central difference scheme or a velocity scheme the displacement $\mathbf{u}^{k+1}$ and velocity $\mathbf{v}^{k+1}$ is obtained.}
	\label{fig:algorithm:bond:chart}
\end{figure}
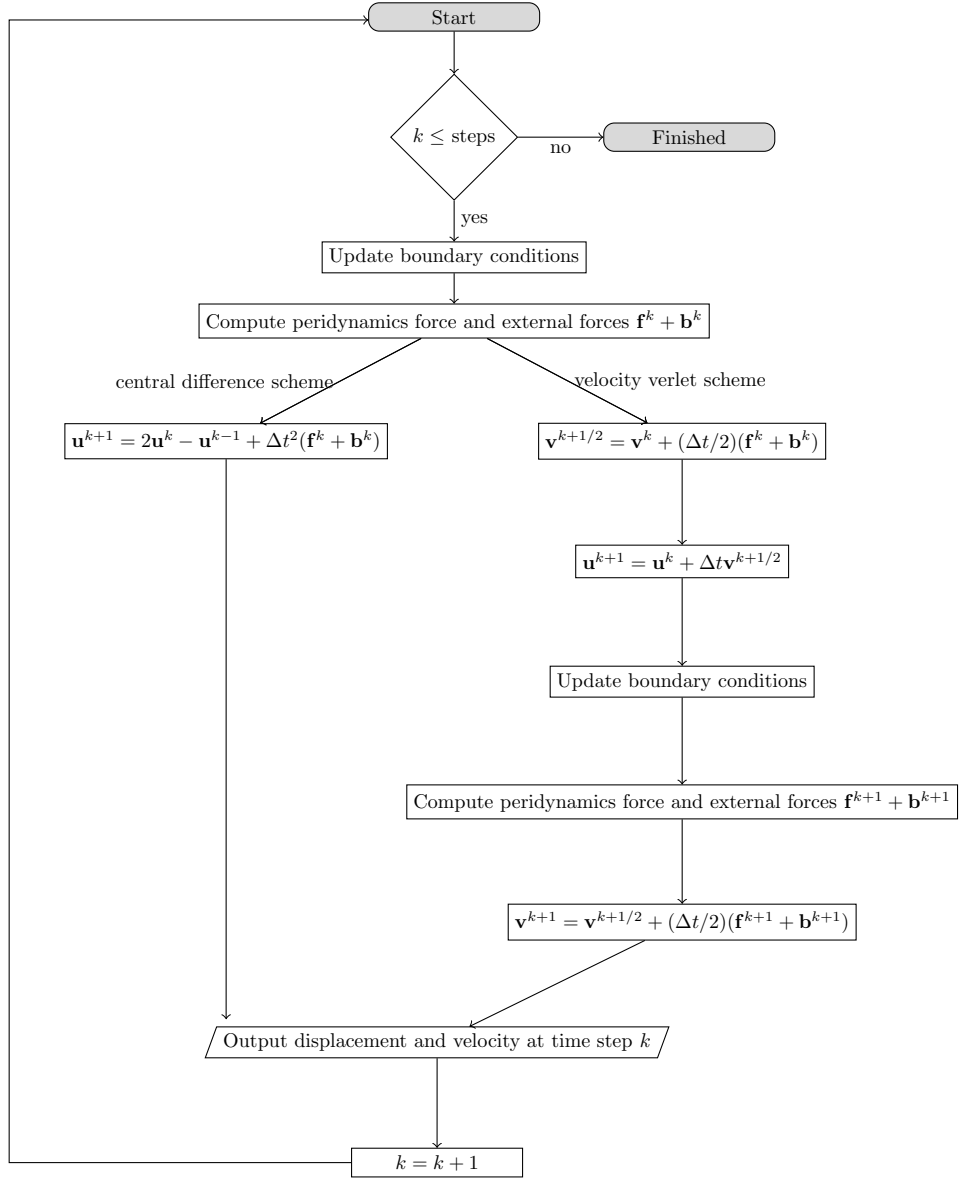

Figure~\ref{fig:algorithm:bond:chart} shows the flow chart for the explicit time integration. 
Algorithm~\ref{alg:time integ verlet} outlines the steps to implement the velocity-verlet scheme
\begin{align}
\mathbf{v}(t^{k+1/2},\mathbf{X}_i) &= \mathbf{v}(t^{k},\mathbf{X}_i) + \frac{(\Delta t/2) }{\varrho(\mathbf{X}_i) } \left[ \mathbf{b}(t^k, \mathbf{X}_i) + \sum\limits_{B_\delta(\mathbf{X}_i)} \mathbf{f}(t^k, \pmb{\eta}, \pmb{\xi}) \right], \label{eq:velocity:verlet:1}\\
\mathbf{u}(t^{k+1},\mathbf{X}_i) &= \mathbf{u}(t^{k},\mathbf{X}_i) + \Delta t \mathbf{v}(t^{k+1/2},\mathbf{X}_i), \label{eq:velocity:verlet:2}\\
\mathbf{v}(t^{k+1},\mathbf{X}_i) &= \mathbf{v}(t^{k+1/2},\mathbf{X}_i) + \frac{(\Delta t/2) }{\varrho(\mathbf{X}_i) } \left[ \mathbf{b}(t^{k+1}, \mathbf{X}_i) + \sum\limits_{B_\delta(\mathbf{X}_i)} \mathbf{f}(t^{k+1}, \pmb{\eta}, \pmb{\xi}) \right]
\label{eq:velocity:verlet}
\end{align}
to obtain the displacement $\mathbf{u}^{k+1}$ for the time step $k+1$. Line~\ref{alg:compute peri force} calls a function of either bond-based Material or state-based Material class to compute the forces and energies corresponding to displacement $\mathbf{u}^k$. The velocity-verlet algorithm is used to compute the velocity at $k+1/2$ and displacement $\mathbf{u}^{k+1}$. Line~\ref{alg:compute peri force 1} invokes the Material class again to compute the forces at new displacements $\mathbf{u}^{k+1}$. The velocity at $k+1$ is computed with the updated forces. The central different scheme is given by 
\begin{align}
\mathbf{u}(t^{k+1},\mathbf{X}_i) = 2\mathbf{u}(t^{k},\mathbf{X}_i) - \mathbf{u}(t^{k-1},\mathbf{X}_i) + \frac{\Delta t^2}{\varrho(\mathbf{X}_i) } \left[ \mathbf{b}(t^k, \mathbf{X}_i) + \sum\limits_{B_\delta(\mathbf{X}_i)} \mathbf{f}(t^k, \pmb{\eta}, \pmb{\xi}) \right]\text{.}
\label{eq:finite:difference}
\end{align}

\begin{algorithm}[tbp]
\caption{Time integration using velocity-verlet scheme}\label{alg:time integ verlet}

\begin{algorithmic}[1]
\State \Comment{Loop over time steps}
\For{$0\leq k < N$}
	\State \Comment{Compute peridynamic force $\mathbf{f}^k$, body force $\mathbf{b}^k$,}
	\State \Comment{external force $\mathbf{f}^k_{ext}$, and total energy $U_{total}^k$ as shown in Algorithm~\ref{alg:bondmaterial}}\label{alg:compute peri force}
	
	\HPXLoop{$i<n$}
		\State \Comment{Compute velocity as shown in Equation~\eqref{eq:velocity:verlet:1}}
		\State	$\mathbf{v}^{k+1/2}(\mathbf{X}_i) = \mathbf{v}^{k}(\mathbf{X}_i) + (\Delta t/2) (\mathbf{f}^k(\mathbf{X}_i) + \mathbf{b}^k(\mathbf{X}_i) + \mathbf{f}^k_{ext}(\mathbf{X}_i))$
		\State \Comment{Compute displacement as shown in Equation~\eqref{eq:velocity:verlet:2}}
		\State $\mathbf{u}^{k+1}(\mathbf{X}_i) = \mathbf{u}^k(\mathbf{X}_i) + \Delta t \mathbf{v}^{k+1/2}\mathbf{f}(\mathbf{X}_i)$
	\EndHPXLoop
	\State Update boundary condition for time $t = (k+1)\Delta t$
	\State \Comment{Compute $\mathbf{f}^{k+1}$, $\mathbf{b}^{k+1}$, $\mathbf{f}^{k+1}_{ext}$, and $U_{total}^{k+1}$ as shown in Algorithm~\ref{alg:bondmaterial}.}\label{alg:compute peri force 1}
	\HPXLoop{$i<n$}\Comment{Loop over nodes}
		\State \Comment{Compute velocity as shown in Equation~\eqref{eq:velocity:verlet}}
		\State	$\mathbf{v}^{k+1}(\mathbf{X}_i) = \mathbf{v}^{k+1/2}(\mathbf{X}_i) + (\Delta t/2) (\mathbf{f}^{k+1}(\mathbf{X}_i) + \mathbf{b}^{k+1}(\mathbf{X}_i) + \mathbf{f}^{k+1}_{ext}(\mathbf{X}_i))$
	\EndHPXLoop
	\State
\EndFor

\end{algorithmic}

\end{algorithm}

\section{Validation of PeridynamicHPX}
\label{sec:validation}
In this section we demonstrate the convergence of HPX implementations for both implicit and explicit schemes.
\subsection{Implicit}

\subsubsection{One dimensional tensile test}
Consider the simple geometry of Figure 9 for comparing the one dimensional implicit time integration against a classical continuum mechanics (CCM) solution. The node on the left-hand side is clamped and displacement is set to zero. A force $F$ is applied to the node at the right-hand side.\\

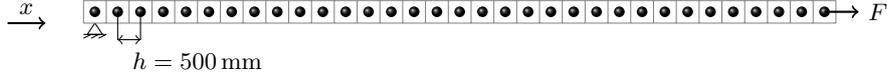
\begin{figure}[tb]
\centering
\begin{tikzpicture}
\draw[->,thick] (-1,0) -- (-0.5,0);
\node[above] at (-0.75,0.) {\small $x$};
\foreach \i in {0,...,32}
{
 \shade[ball color = black](0.15+0.3*\i,0.15) circle (2pt);

}
\draw (0.15,0) -- (0.05,-0.15);
\draw (0.15,0) -- (0.25,-0.15);
\draw (0.05,-0.15) -- (0.25,-0.15);
\draw (0.,-0.15) -- (0.3,-0.15);
\draw (0.2,-0.15) -- (0.1,-0.2);
\draw (0.1,-0.15) -- (0.0,-0.2);
\draw (0.3,-0.15) -- (0.2,-0.2);
%
\draw[step=0.3,gray,very thin] (-0.,0) grid (9.9,0.3);
\draw[->,thick] (9.75,0.15) -- (10.2,0.15);
\node[right] at (10.2,0.15) {\small $F$};
\draw (0.45,0.15) -- (0.45,-0.225);
\draw (0.75,0.15) -- (0.75,-0.225);
\draw[<->] (0.45,-0.22) -- (0.75,-0.22);
\node at (1.5,-0.5) {\small $h=500\,\si{\milli\meter}$};
\end{tikzpicture}
\caption{Sketch of the one dimensional bar benchmark test. The node on the left-hand side is clamped and its displacement is set to zero. A force $F$ is applied on the node at the right-hand side. Adapted from~\cite{Diehl2020_validation_1d}.}
\label{fig:sketch:validation:im1}
\end{figure}
\noindent
The strain, stress, and strain energy for this configuration \yellow{are} compared with the values obtained from classical continuum mechanics (CCM) where $\sigma = E\cdot\varepsilon$, where $\sigma$ is the stress, $E$ is the Young's modulus and $\varepsilon$ is the strain. The stress $\sigma=\sfrac{F}{A}$, is then defined by the force $F$ per cross section $A$. Thus, the strain is obtained by $\varepsilon=\sfrac{\sigma}{E}=\sfrac{F}{(AE)}$. For a force $F$ of $40\,\si{\newton}$, a cross section of $1\,\si{\square\meter}$ and a Young's modulus of $4\,\si{\giga\pascal}$, the resulting strain reads $\varepsilon_\text{CCM}=\num{1e-8}$ an the stress is of $\sigma_\text{CCM}=40\,\si{\pascal}$. The strain energy density is given by $U_\text{CCM}=\sfrac{\sigma^2}{(2E)}=\num{2e-7}\,\si{\pascal}$.\\  

\begin{figure}[ptb]
 \centering
\begin{subfigure}[t]{0.5\textwidth}
       
        \includegraphics[width=\textwidth]{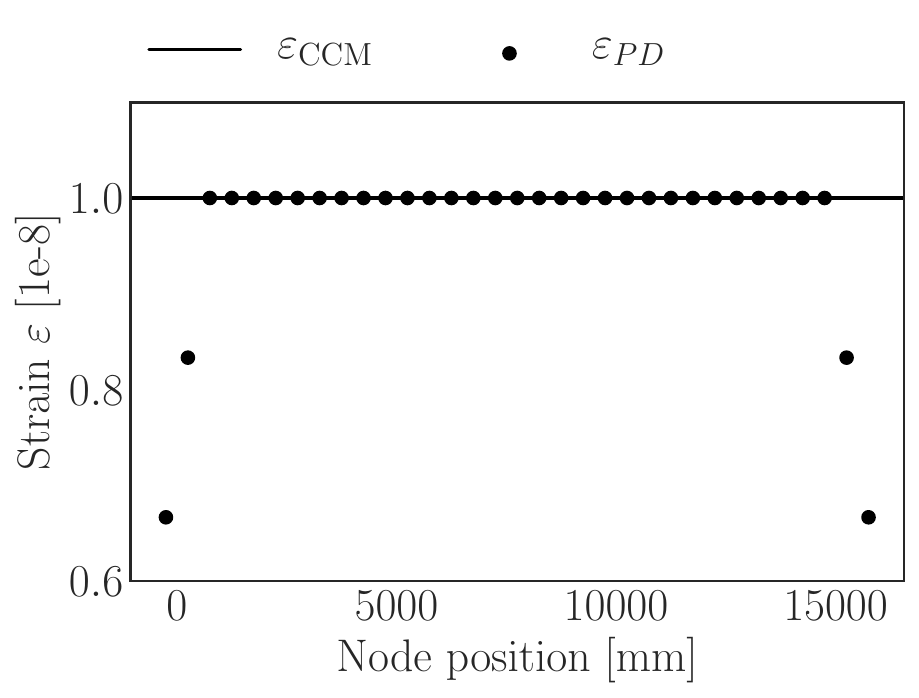}
        \caption{}
        \label{fig:validation:im1:strain}
    \end{subfigure}
    
\begin{subfigure}[t]{0.5\textwidth}
       
        \includegraphics[width=\textwidth]{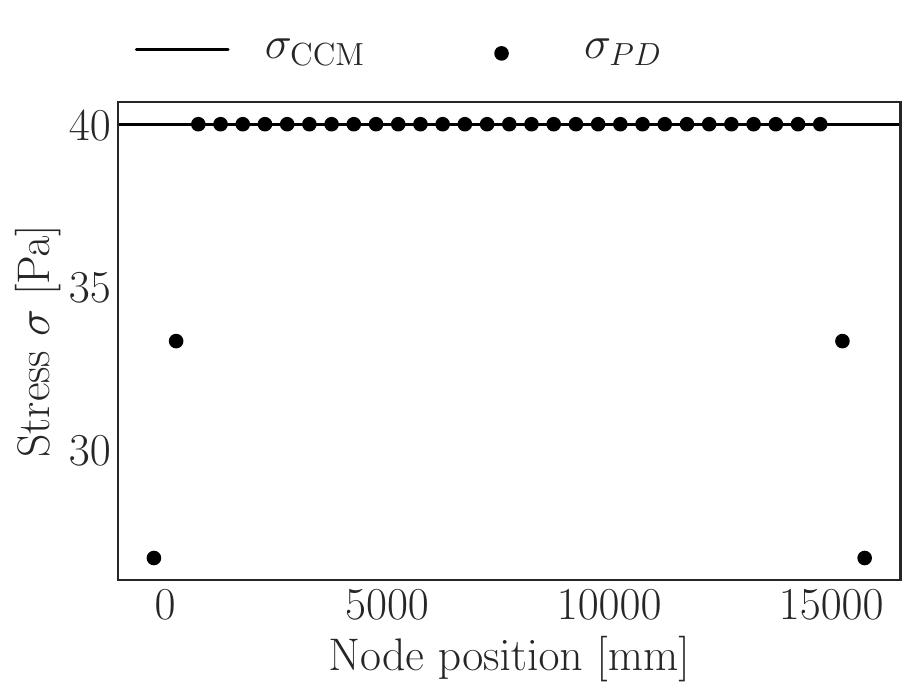}
        \caption{}
        \label{fig:validation:im1:stress}
    \end{subfigure}
    
\begin{subfigure}[t]{0.5\textwidth}
        \includegraphics[width=\textwidth]{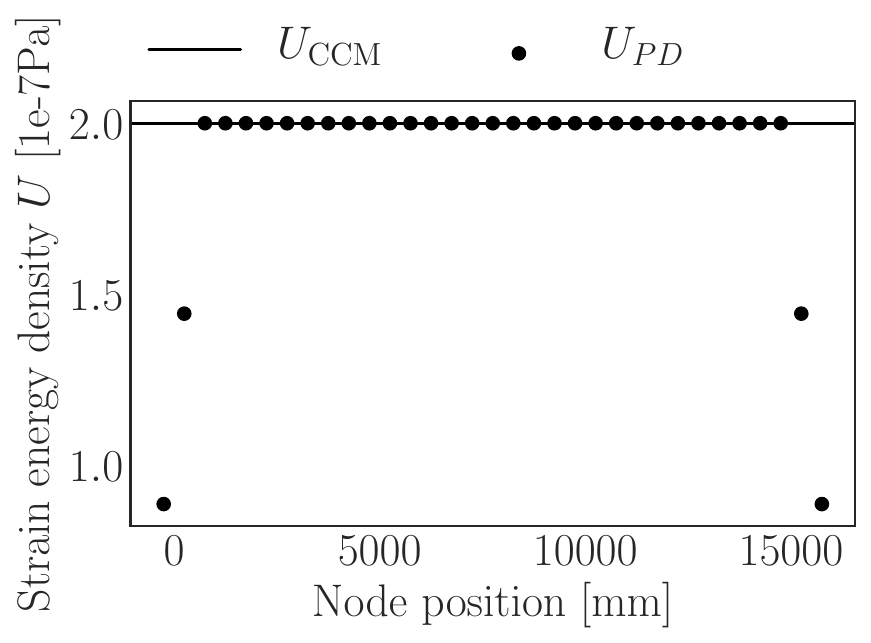}
        \caption{}
        \label{fig:validation:im1:energy}
    \end{subfigure}
\caption{Comparison of strain $\epsilon$ (\subref{fig:validation:im1:strain}), stress $\sigma$ (\subref{fig:validation:im1:stress}), and strain energy $U$ (\subref{fig:validation:im1:energy}) with classical continuum mechanics. Close to the boundary the surface effect influences the accuracy.}
\label{fig:validation:im1}
\end{figure}
\noindent
The bar was discretized with $33$ nodes with a nodal spacing $h$ of $500\,\si{\milli\meter}$. The horizon was $\delta=1000\,\si{\milli\meter}$. The tolerance $\tau$ for the Biconjugate gradient stabilized method (BiCGSTAB) was \num{1e-9}. Figure~\ref{fig:validation:im1} shows that stresses, strains and strain energy match perfectly the theoretical values inside the bar but all these quantities diverge at the boundaries. These effects are the well-known surface effects within the EMU nodal discretization.
\subsubsection{Two dimensional tensile test}
\label{sec:validation:im:2}
\begin{figure}[tb]
\centering
\begin{tikzpicture}
\draw[step=0.5,gray,very thin] (-0.5,0) grid (3,3.5);
\draw[->,thick] (-2.5,0) -- (-2,0);
\draw[->,thick] (-2.5,0) -- (-2.5,0.5);
\node[above] at (-2.5,0.5) {\small $y$};
\node[right] at (-2.,0.) {\small $x$};
\foreach \i in {-1,...,5}
\foreach \j in {0,1,2,4,5,6}
\shade[ball color = black](0.25+\i*0.5,0.25+\j*0.5) circle (2pt);
\foreach \i in {-1,0,1,3,4,5}
\foreach \j in {3}
\shade[ball color = black](0.25+\i*0.5,0.25+\j*0.5) circle (2pt);
\shade[ball color = black](0.25+1.,0.25+1.5) circle (2pt);

\foreach \i in {0,...,6}
\draw[->,thick] (-0.25,0.25+\i*0.5) -- (-0.75,0.25+\i*0.5);
\node[left] at (-0.75,1.75) {\small $F$ };

\draw (-0.5,0) -- (-0.5,-0.35);
\draw (3,0) -- (3,-0.35);
\draw[<->] (-0.5,-0.3) -- (3,-0.3);
\node[below] at (1.25,-0.35) {\small $W=375\,\si{\milli\meter}$};
\draw (-1.35,0) -- (0,0);
\draw (-1.35,3.5) -- (0.0,3.5);
\draw[<->] (-1.3,0) -- (-1.3,3.5);
\node[above,rotate=90] at (-1.3,1.75) {\small $H=375\,\si{\milli\meter}$};

\node[above] at (2.75,4.) {\small $h=25\,\si{\milli\meter}$};
\draw[<->] (2.5,3.95) -- (3.,3.95); 
\draw (3,3.5) -- (3,4);
\draw (2.5,3.5) -- (2.5,4);

\foreach \i in {0,...,6}
{
\draw (2.78,0.25+\i*0.5) -- (3.15-0.25,0.1+\i*0.5);
\draw (2.78,0.25+\i*0.5) -- (3.15-0.25,0.4+\i*0.5);
\draw (3.15-0.25,0.4+\i*0.5) -- (3.15-0.25,0.1+\i*0.5);
\draw (3.25-0.35,0.03+\i*0.5) -- (3.25-0.35,0.45+\i*0.5);
\draw (3.25-0.35,0.4+\i*0.5) -- (3.3-0.35,0.3+\i*0.5);
\draw (3.25-0.35,0.3+\i*0.5) -- (3.3-0.35,0.2+\i*0.5);
\draw (3.25-0.35,0.2+\i*0.5) -- (3.3-0.35,0.1+\i*0.5);
}

\foreach \i in {0,...,6}
{
\draw (0.25+2.5,0.24+\i*0.5) -- (0.1+2.5,-0.15+0.25+\i*0.5);
\draw (0.25+2.5,0.24+\i*0.5) -- (0.4+2.5,-0.15+0.25+\i*0.5);
\draw (0.1+2.5,-0.15+0.25+\i*0.5) -- (0.4+2.5,-0.15+0.25+\i*0.5);
\draw (0.03+2.5,-0.25+\i*0.5+0.35) -- (0.45+2.5,-0.25+\i*0.5+0.35);
\draw (0.4+2.5,-0.25+0.35+\i*0.5) -- (0.3+2.5,-0.3+0.35+\i*0.5);
\draw (0.3+2.5,-0.25+0.35+\i*0.5) -- (0.2+2.5,-0.3+0.35+\i*0.5);
\draw (0.2+2.5,-0.25+0.35+\i*0.5) -- (0.1+2.5,-0.3+0.35+\i*0.5);
}
\end{tikzpicture}
\caption{Sketch of the two dimensional tensile test. All nodes on the line of the right-hand side of the plate are clamped in $x$-direction and $y$-direction. A force of $-50$\,\si{\kilo\newton} is applied in $x-direction$ to each node of the line on the left-hand side. Adapted from~\cite{diehl_2020_2d}.}
\label{fig:sketch:validation:im2}
\end{figure}
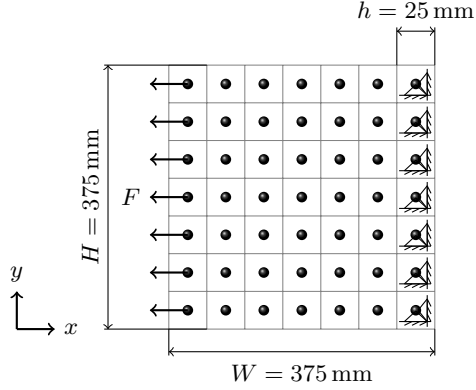

Figure~\ref{fig:sketch:validation:im2} shows the two-dimensional tensile benchmark. The line of nodes of the right-hand side  of the plate are clamped in $x$-direction and $y$-direction. On each of the nodes of the line at the left-hand side a force of force $F=-50$\,\si{\kilo\newton} in $x$-direction was applied. The displacement of a node $\mathbf{X}_i$ for a tensile behavior can be derived using the Airy stress function~\cite{sadd2009elasticity} as follows
\begin{align}
\mathbf{u}_x(\mathbf{X}_i) &= \frac{F}{EWT} (\mathbf{X}_{i_x}-W)  \text{,}\label{eq:ccm:2D} \\\nonumber
\mathbf{u}_y(\mathbf{X}_i) &= -\frac{\nu F}{E T}\left( \frac{\mathbf{X}_{i_y}}{W}  - \frac{1}{2}\right)\text{,} \quad
\end{align}
where $F$ is the applied force and $W$ and $H$ and $T$ are respectively the plate's width and height and thickness. Note that we assume a thickness $T=1,\si{\milli\meter}$. $\mathbf{X}_{i_x}, \mathbf{X}_{i_y}$ denotes the x and y coordinate of node $\mathbf{X}_i$. A Young's modulus of $E=4000$\,\si{\mega\pascal} and a Poisson's ratio of $\varrho=0.3$ was used.\\

$H$ and $W$ were set to $375$\,\si{\milli\meter} and $h=25$\,\si{\milli\meter}. The tolerance for the BiCGSTAB solver was $\tau=\num{1e-3}$. The $m_d$-value was, $4$, which means that $2m_d+1$ nodes are within $[\mathbf{X}_{i_x}-\delta,\mathbf{X}_{i_x}+\delta]$ and $[\mathbf{X}_{i_y}-\delta,\mathbf{X}_{i_y}+\delta]$.
Table~\ref{tab:validation:implicit:2D} lists the actual position at the node in the center of the plate $x_\text{mid}$ and its comparison with the one from CCM~\eqref{eq:ccm:2D}. The relative error for the actual position in $x$-direction is sufficiently small. With the applied boundary conditions the displacement at the point in the center of the plate in $y$-direction is zero.

\begin{table}[tp]
\centering
\begin{tabular}{llll}
\toprule
Actual position & CCM & PD & Relative error \\\midrule
$x$-direction & \num{0.18749}\,\si{\si{\meter}} & \num{0.18744}\,\si{\si{\meter}} & \num[scientific-notation=true]{4.9e-05}  \\
$y$-direction & \num{0.1875}\,\si{\si{\meter}} & \num{0.1875}\,\si{\si{\meter}} & \num{0} \\
\bottomrule
\end{tabular}
\caption{Comparison of the actual position in meters of the node in the center of the plate obtained in the simulation with those from classical continuum mechanics~\eqref{eq:ccm:2D}.}
\label{tab:validation:implicit:2D}
\end{table}

\subsection{Explicit}
\label{sec:validation:ex:2}
For the explicit scheme, we present numerical verification of convergence of the approximation. We study the rate at which numerical approximation converge with respect to the mesh size. Similar to [Section 6.1.2, \cite{CMPer-JhaLipton5}], we derive the formula to compute the rate of convergence numerically. We denote the $L^2$ norm of function $\mathbf{f} : \Omega_0 \to \R^d$ as $||\mathbf{f}|| := \sqrt{\int_{\Omega_0} |\mathbf{f}(\mathbf{X})|^2 d\mathbf{X}}$, where $d\mathbf{X}$ is the infinitesimal length (1-d), area (2-d), or volume (3-d) element for given dimension $d$. 

Let $\mathbf{u}_1, \mathbf{u}_2, \mathbf{u}_3$ are three approximate displacement fields corresponding to the meshes of size $h_1, h_2, h_3$. Let $\mathbf{u}$ be the exact solution. We assume that for $h' < h$, $\underline{C}h^\alpha \leq ||\mathbf{u}_h - \mathbf{u}_{h'}|| = \bar{C} h^\alpha$ with $\underline{C} \leq \bar{C}$ and $\alpha > 0$. We fix the ratio of mesh sizes as $\frac{h_1}{h_2} = \frac{h_2}{h_3} = r$, where $r$ is fixed number. We can show then
\begin{align}
\alpha \leq \frac{\log(||\mathbf{u}_1 - \mathbf{u}_2||) - \log(||\mathbf{u}_2 - \mathbf{u}_2||) + \log(\bar{C}) - \log(\underline{C})}{\log(r)}.
\end{align}
So an upper bound on the convergence rate is at least as big as
\begin{align}\label{eq:rate lower bound}
\bar{\alpha} = \frac{\log(||\mathbf{u}_1 - \mathbf{u}_2||) - \log(||\mathbf{u}_2 - \mathbf{u}_3||)}{\log(r)}.
\end{align}
$\bar{\alpha}$ provides an estimate for rate of convergence. 


We now present the convergence results for explicit scheme. 
\subsubsection{One dimensional}\label{sec:validation:exp:1D}
\noindent
In 1-d the strain between material point $X,X'$ is given by $S(X',X) = \frac{u(X') - u(X)}{|X' - X|}$. We consider a nonlinear peridynamic force between material point $X,X'$ of the form, see [Section 2.1, \cite{CMPer-JhaLipton2}],
\begin{align}\label{eq:per force nonlinear}
f(t,&u(t,X')-u(t,X), X'-X))= \notag\\ & \dfrac{2}{\delta^2} \int_{X-\delta}^{X+\delta} J^\delta(|X' - X|) \psi'(|X' - X| S(X',X)^2) S(X',X) dX'\text{.}
\end{align}
$\psi: \R^+ \to \R$ is the nonlinear potential which is smooth, positive, and concave. We set function $\psi$ as
\begin{align}\label{eq:function_psi}
\psi(r) = C(1 - \exp[-\beta r]), \qquad C = \beta = 1.
\end{align}
With this choice of potential, we effectively model a bond which is elastic for small strains and softens for large strains. As $S\to \infty$, $\psi'(r) = C\beta \exp[-\beta r] \to 0$, and therefore, the pairwise force $f$ between two points $X,X'$ go to zero as the bond-strain $S(X',X)$ gets larger and larger. The cumulative effect is that the material behaves like a elastic material under small deformation and cracks develop in regions where the deformation is large, see \cite{CMPer-Lipton2}. 
The influence function $J^\delta$ is of the form: $J^\delta(r) = J(r/\delta)$, where $J(r) = c_1 r \exp(-r^2/c_2)$, $c_1 = 1$, and $c_2 = 0.4$. \\
The linear peridynamic force is given by, see [Section 2.1, \cite{CMPer-JhaLipton2}],
\begin{align}\label{eq:per force linear}
f_l(t,&u(t,X')-u(t,X), X'-X))= \notag\\ & \dfrac{2}{\delta^2} \int_{X-\delta}^{X+\delta} J^\delta(|X' - X|) \psi'(0) S(X',X) dX'\text{.}
\end{align}
From \eqref{eq:function_psi}, we have $\psi'(0) = C \beta$.

\begin{figure}[tbp]
\centering
\includegraphics[width=0.85\linewidth]{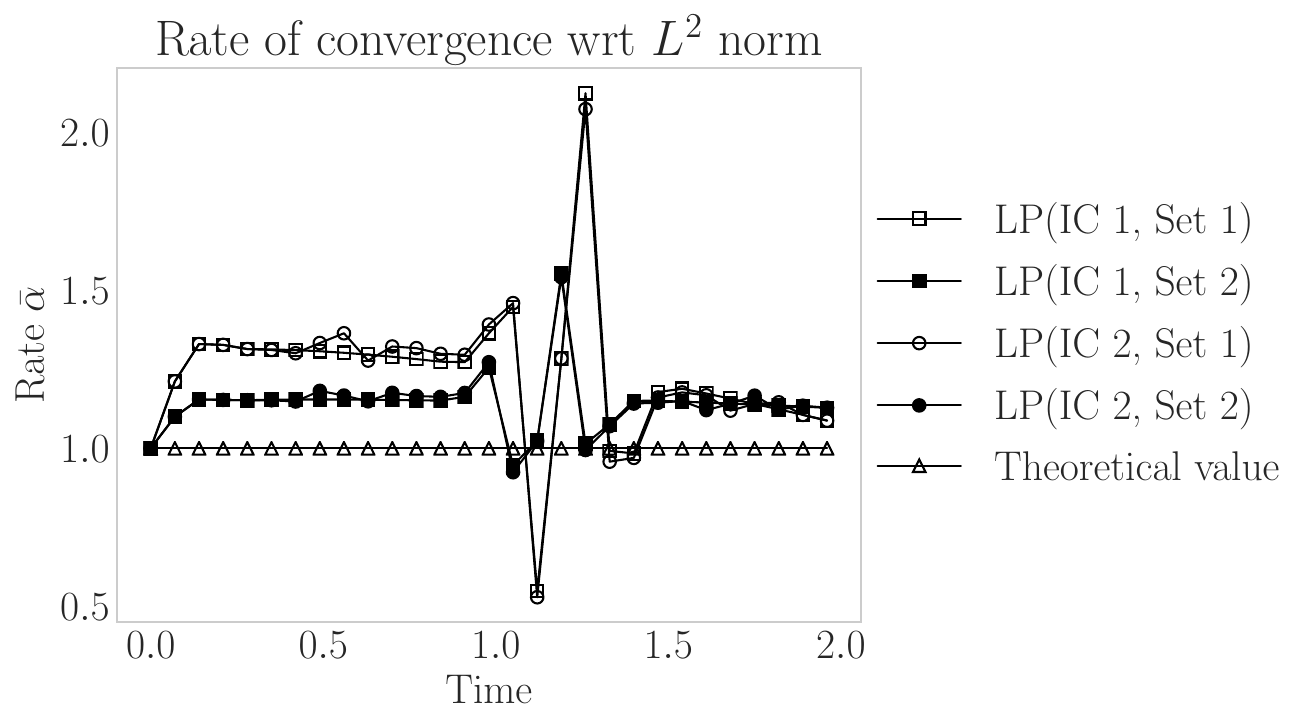}
\caption{Time vs rate of convergence with respect to mesh size. These results are for linear peridynamic force (LP). Set 1 corresponds to convergence rate obtained from solutions of mesh sizes $\{h_1,h_2,h_3\}$ and set 2 corresponds to convergence rate obtained from solutions of mesh sizes $\{h_2,h_3,h_4\}$. The boundary condition is $u=0$ on non-local boundary $\Omega_c = [-\delta, 0] \cup [1,1+\delta]$. These results validate the implementation of the explicit scheme for peridynamics in one dimension.}
\label{fig:rate convergence 1d dirichlet form}
\end{figure}

Let $\Omega = [0,1]$ be the material domain with an horizon $\delta = 0.01$. The time domain is $[0,2]$ with a time step $\Delta t = 10^{-5}$. 
Consider four mesh sizes $h_1=\delta/2$, $h_2=\delta/4$, $h_3=\delta/8$, and $h_4=\delta/16$, and compute equation~\eqref{eq:rate lower bound} for two sets $\{h_1,h_2,h_3\}$ and $\{h_2,h_3,h_4\}$ of mesh sizes. The boundary conditions are those described in Figure~\ref{fig:rate convergence 1d dirichlet form}. Apply either one of the initial conditions:\\

\textbf{Initial condition 1(IC 1): }Let the initial condition on the displacement $u_0$ and the velocity $v_0$ be given by
\begin{align}
u_0(X) = \exp[-|X - x_c|^2/\beta] a, \quad v_0(X) = 0,
\end{align}
with $x_c = 0.5, a= 0.001, \beta = 0.003$. $u_0$ is the Gaussian function centered at $x_c$.\\

\textbf{Initial condition 2(IC 2): }The initial condition $u_0$ and $v_0$ are described as
\begin{align}
u_0(X) = \exp[-|X - x_{c1}|^2/\beta] a + \exp[-|X - x_{c2}|^2/\beta] a, \quad v_0(X) = 0,
\end{align}
with $x_{c1} = 0.25, x_{c2} = 0.75, a= 0.001, \beta = 0.003$. $u_0$ is the sum of two Gaussian functions centered at $x_{c1}$ and $x_{c2}$.\\

\begin{figure}[tbp]
\centering
\includegraphics[width=0.85\linewidth]{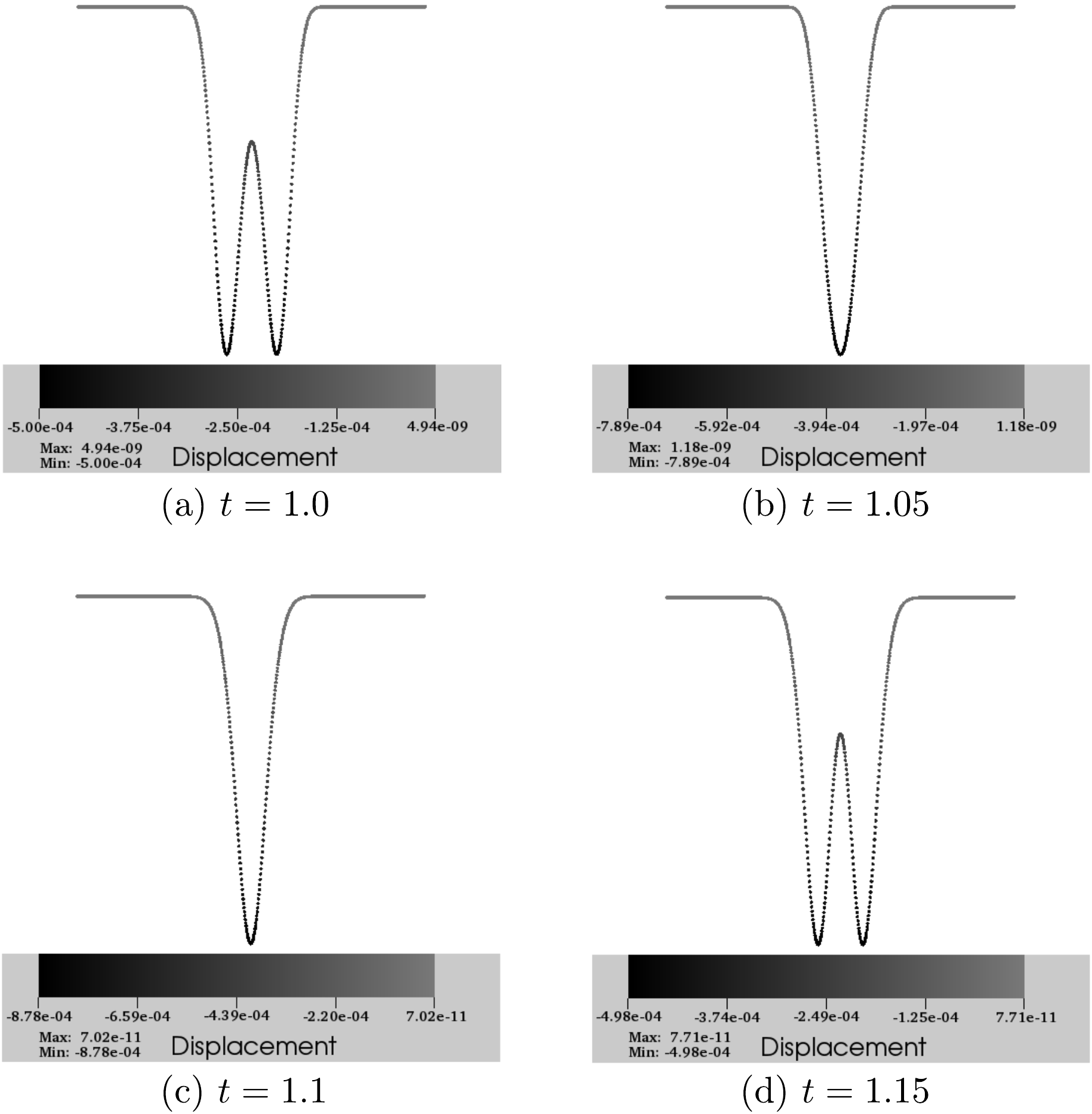}
\caption{Displacement profile for mesh size $h = \delta /8$. Results are for IC 1 and for linear peridynamic (LP). In (a) the waves are traveling towards each other. In (b) and (c) we see increase in amplitude after merging. In (d) two waves are traveling away from each other.}
\label{fig:displacement_1d_ic1}
\end{figure}

\begin{figure}[tbp]
\centering
\includegraphics[width=0.85\linewidth]{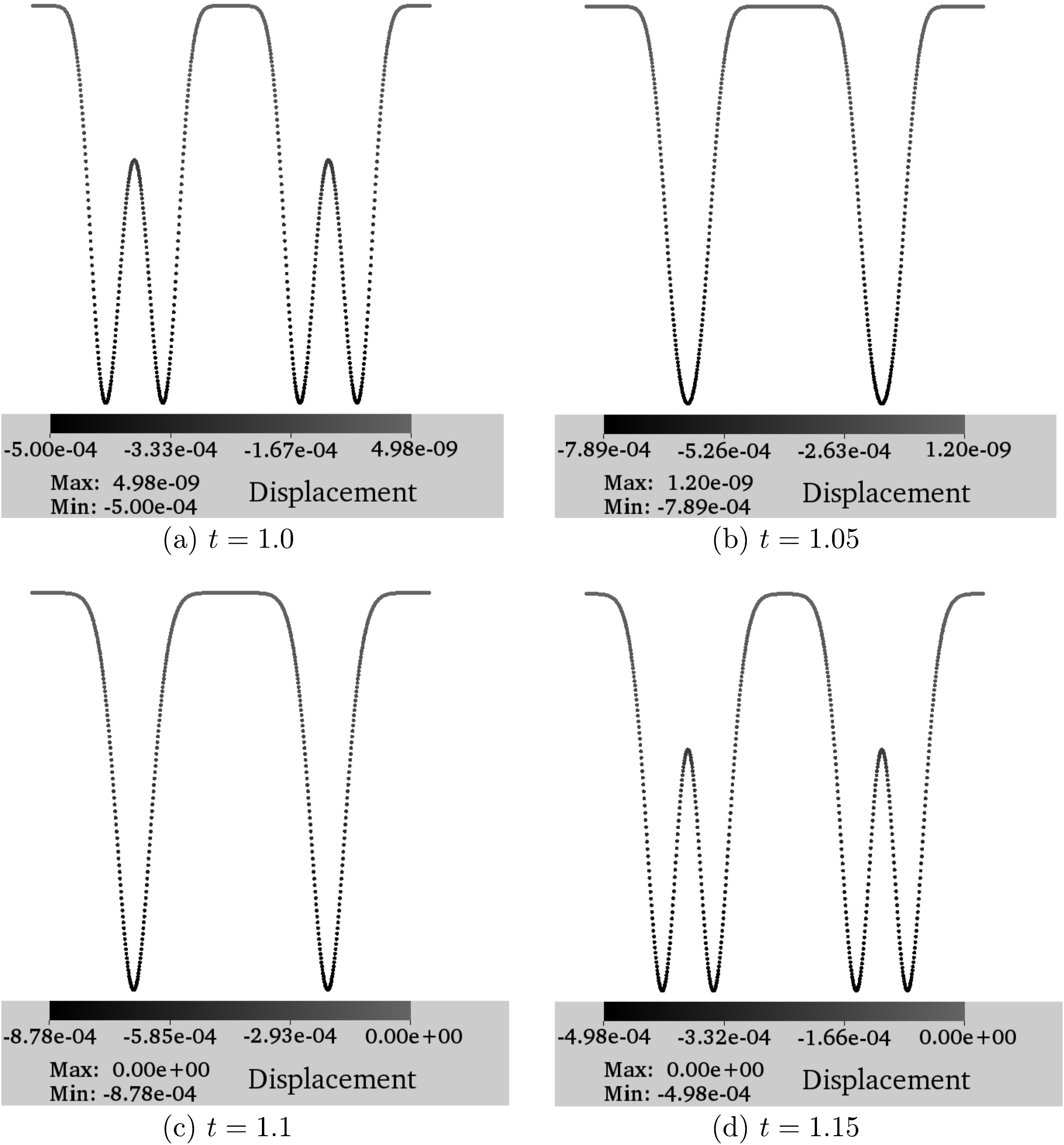}
\caption{Displacement profile for mesh size $h = \delta /8$. Results are for IC 2 and for linear peridynamic (LP). In (a) the two waves in left side (X=0.25) and right side (X=0.75) approach towards each other. (b) and (c) correspond to intermediate time before the wave divides into two smaller waves moving in opposite direction in (d). }
\label{fig:displacement_1d_ic2}
\end{figure}

Figure~\ref{fig:rate convergence 1d dirichlet form} shows the rate of convergence $\bar{\alpha}$ as a function of time for solutions having the initial conditions 1 and 2. The convergence rate for $\{h_1,h_2,h_3\}$ and $\{h_2,h_3,h_4\}$ follows the same trend.  The bump in the Figure~\ref{fig:rate convergence 1d dirichlet form} near $t=1.1$ for both the initial conditions is due to the merging of two waves traveling towards each other. We show the displacement profile near $t=1.1$ in Figure~\ref{fig:displacement_1d_ic1} for IC 1 and Figure~\ref{fig:displacement_1d_ic2} for IC 2. 
This behavior is particularly difficult to capture and requires much finer mesh. 
For the rapidly varying (spatially) displacement field, the length scale at which the displacement varies is small, and requires very fine mesh size. 
This is the reason we see a better rate for Set 2 $\{h_2, h_3, h_4 \}$ as compared to Set 1 $\{h_1, h_2, h_3 \}$ in Figure~\ref{fig:rate convergence 1d dirichlet form}. This behavior is true for both the initial conditions.
Similar results, not shown here, were obtained for the nonlinear model of equation~\eqref{eq:motion:bond}. The convergence results presented in Figure~\ref{fig:rate convergence 1d dirichlet form} agree with the theoretical convergence rate, which suggests that the implementation is robust.

\subsubsection{Two dimensional}
\label{sec:validation:exp:2D}
In higher dimension the strain is given by $S(\mathbf{X}',\mathbf{X}) = \frac{\mathbf{u}(\mathbf{X}') - \mathbf{u}(\mathbf{X})}{|\mathbf{X}' - \mathbf{X}|} \cdot \frac{\mathbf{X}' - \mathbf{X}}{|\mathbf{X}' - \mathbf{X}|}$. The nonlinear peridynamic force in 2-d is given by, see \cite{CMPer-Lipton},
\begin{align}\label{eq:per force nonlinear2d}
\mathbf{f}(t,&\mathbf{u}(t,\mathbf{X}')-\mathbf{u}(t,\mathbf{X}), \mathbf{X}'-\mathbf{X}))= \notag\\ & \dfrac{4}{\delta |B_\delta(\mathbf{0})|} \int_{B_\delta(\mathbf{X})} J^\delta(|\mathbf{X}' - \mathbf{X}|) \psi'(|\mathbf{X}' - \mathbf{X}| S(\mathbf{X}',\mathbf{X})^2) S(\mathbf{X}',\mathbf{X})\frac{\mathbf{X}' - \mathbf{X}}{|\mathbf{X}' - \mathbf{X}|} d\mathbf{X}'\text{.}
\end{align}
$\psi(r)$ is given by \eqref{eq:function_psi}. We set $J^\delta(r) = 1$ if $r< 1$ and $0$ otherwise. Similar to the 1-d case, if we substitute $\psi'(0)$ in place of $\psi'(|\mathbf{X}' - \mathbf{X}| S(\mathbf{X}',\mathbf{X})^2)$ we will get the expression of linear peridynamic force $\mathbf{f}_l$. 

Let $\Omega = [0,1]^2$ be the material domain with a horizon $\delta = 0.1$.   
The 2-d vector $\mathbf{X}$ is written $\mathbf{X} = (X_1,X_2)$ where $X_1$ and $X_2$ are the components along the $x$ and $y$ axes. 
The time domain is taken to be $[0,2]$ and $\Delta t = 10^{-5}$ is the time step. The  influence function is $J^\delta(r) = 1$ for $r<\delta$ and $0$ otherwise. The rate $\bar{\alpha}$ is computed for three mesh sizes $h = \delta/2, \delta/4, \delta/8$. 

\begin{figure}[tbp]
\centering
\includegraphics[width=0.3\linewidth]{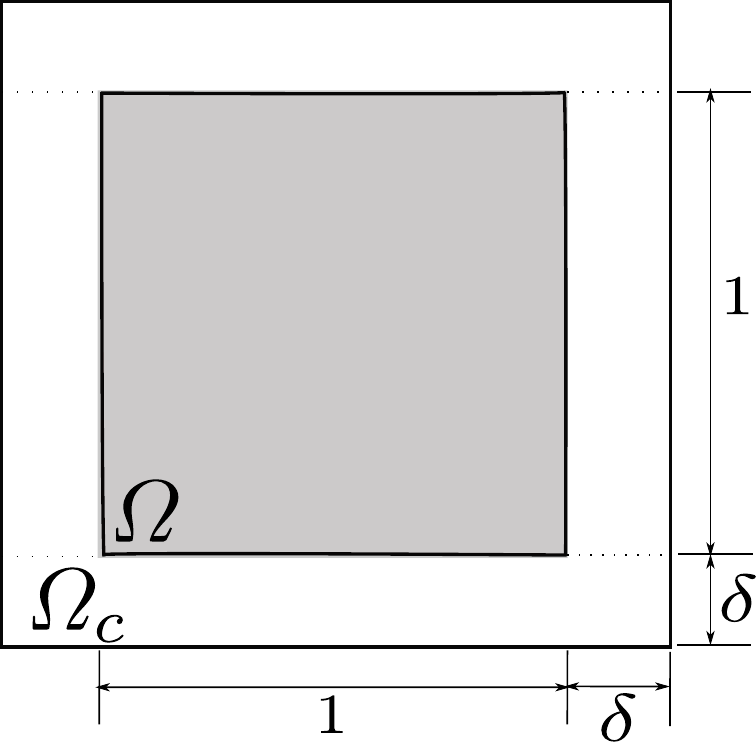}
\caption{Square domain (grey area) $\Omega = [0,1]^2$ and a nonlocal boundary  $\Omega_c = [-\delta, 1+\delta]^2-[0,1]^2$. The area outside $\Omega$ and within the outer boundary is $\Omega_c$. Dashed lines show the division of $\Omega_c$ into left, right, bottom, and top parts.}
\label{fig:nonlocal boundary}
\end{figure}

We fix $\mathbf{u} = (0,0)$ on the collar $\Omega_c$ around domain $\Omega$, see Figure~\ref{fig:nonlocal boundary}. Let the initial condition on displacement vector $\mathbf{u}_0 = (u_{0,1}, u_{0,2})$ and velocity vector $\mathbf{v}_0 = (v_{0,1}, v_{0,2})$ be
\begin{align}\label{eq:ic-2d}
u_{0,1}(X_1, X_2) &= \exp[-(|X_1 - x_{c,1}|^2 + |X_2 - x_{c,2}|^2) /\beta] a_1, \notag \\
u_{0,2}(X_1, X_2) &= \exp[-(|X_1 - x_{c,1}|^2 + |X_2 - x_{c,2}|^2) /\beta] a_2, \notag \\
v_{0,1}(X_1, X_2) &= 0, v_{0,2}(X_1, X_2) = 0,
\end{align}
where $\mathbf{a} = (a_1,a_2)$ and $\mathbf{x}_c = (x_{c,1}, x_{c,2})$ are 2-d vectors and $\beta$ is a scalar parameter. Two different types of initial conditions are considered: 

\textbf{Initial condition 1(IC 1): } 
\begin{align}\label{eq:ic-2d-1}
\mathbf{a} = (0.001, 0), \mathbf{x}_c = (0.5, 0.5), \text{ and } \beta = 0.003\text{.}
\end{align}

\textbf{Initial condition 2(IC 2): }
\begin{align}\label{eq:ic-2d-2}
\mathbf{a} = (0.0002, 0.0007), \mathbf{x}_c = (0.25, 0.25), \text{ and } \beta = 0.01\text{.} 
\end{align}

\begin{figure}[tbp]
\centering
\includegraphics[width=0.85\linewidth]{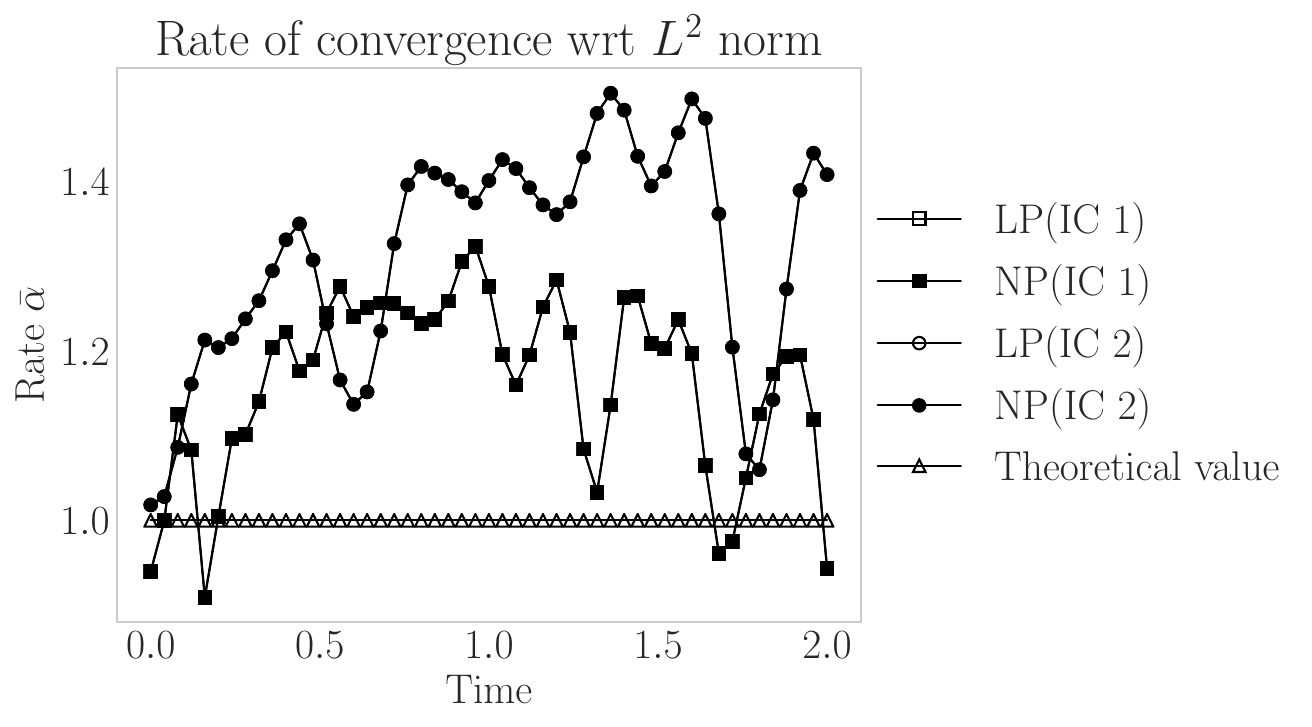}
\caption{Time vs rate of convergence with respect to mesh size. The boundary condition is $\mathbf{u}=(0,0)$ on the non-local boundary $\Omega_c = [-\delta, 1+\delta]^2 - [0,1]^2$. IC 1 and IC 2 refer to the two initial conditions described in Equation~\eqref{eq:ic-2d-1} and Equation~\eqref{eq:ic-2d-2}. The rate of convergence is similar for linear (LP) and nonlinear (NP) peridynamics. 
}
\label{fig:rate convergence 2d dirichlet form}
\end{figure}

Figure~\ref{fig:rate convergence 2d dirichlet form} shows $\bar{\alpha}$ with respect to time for the nonlinear (NP) and linear (LP) peridynamics solutions. Solutions appear to converge at a rate above theoretically predicted rate of $1$ for the nonlinear model with boundary conditions considered in this problem, see~\cite{CMPer-JhaLipton,CMPer-JhaLipton2}.

\section{Benchmark for PeridynamicHPX}
\label{sec:benchmark}

\subsection{Implicit}
%

\begin{figure}[tbp]
        \centering
        \includegraphics[width=0.6\textwidth]{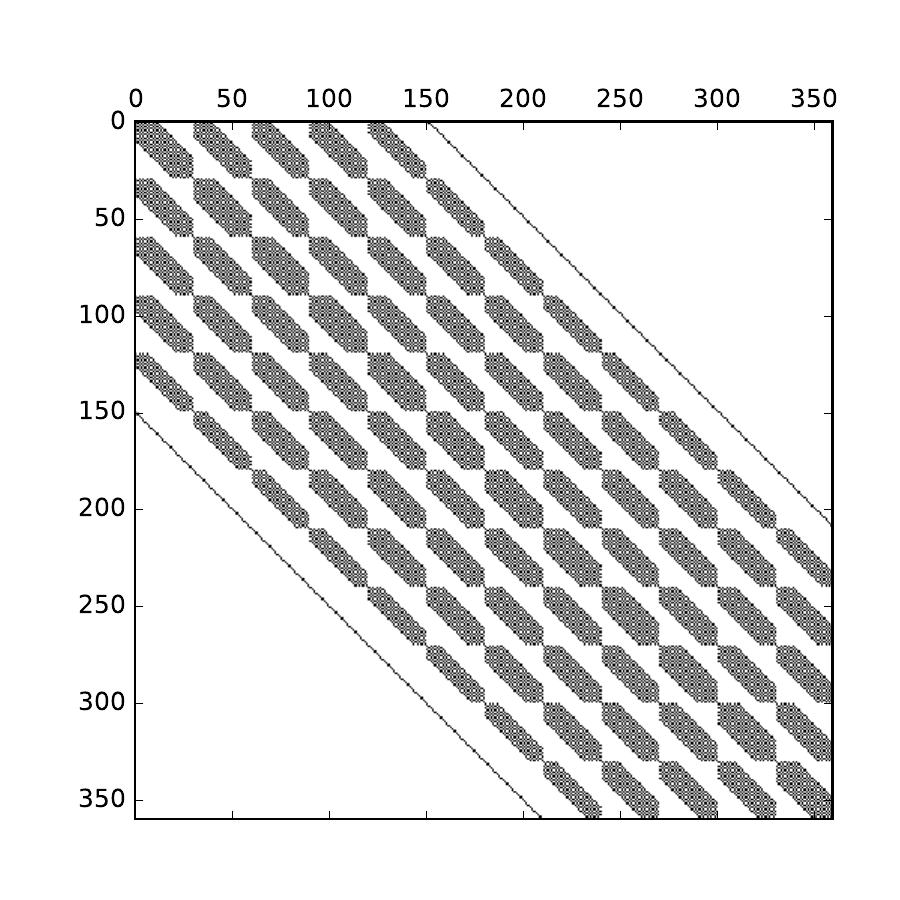}
        \caption{Nonzero matrix elements $(20436)$ of the tangent stiffness matrix $K$ with the condition number $\kappa(K)=\sfrac{\vert K \vert_{L_2}}{\vert K^{-1}\vert}_{L_2}=90.688$ as defined in~\cite{strang1980}.}
        \label{fig:benchmark:im2:matrix}
\end{figure}

The test case of Section~\ref{sec:validation:im:2} served as benchmark for the two dimensional implicit time integration. Figure~\ref{fig:benchmark:im2:matrix} shows the $20436$ nonzero entries of the tangent stiffness matrix $K^{360\times 360}$ with the condition number $\kappa(K)=90.688$. The solver required $28$ iterations. The benchmark was run on Fedora $25$ with kernel $4$.$8$.$10$ on Intel(R) Xeon(R) CPU E$5$-$1650$ v$4$ @ $3$.$60$GHz with up to $6$ physical cores. HPX (version $bd2f240$ $b1565b4$) and PeridynamicHPX were compiled with gcc $6.2.1$, boost $1$.$61$ and blaze $3$.$2$ libraries were used.\\ 


\begin{figure}[tbp]
\centering
\begin{subfigure}[t]{0.6\textwidth}
       \includegraphics[width=\textwidth]{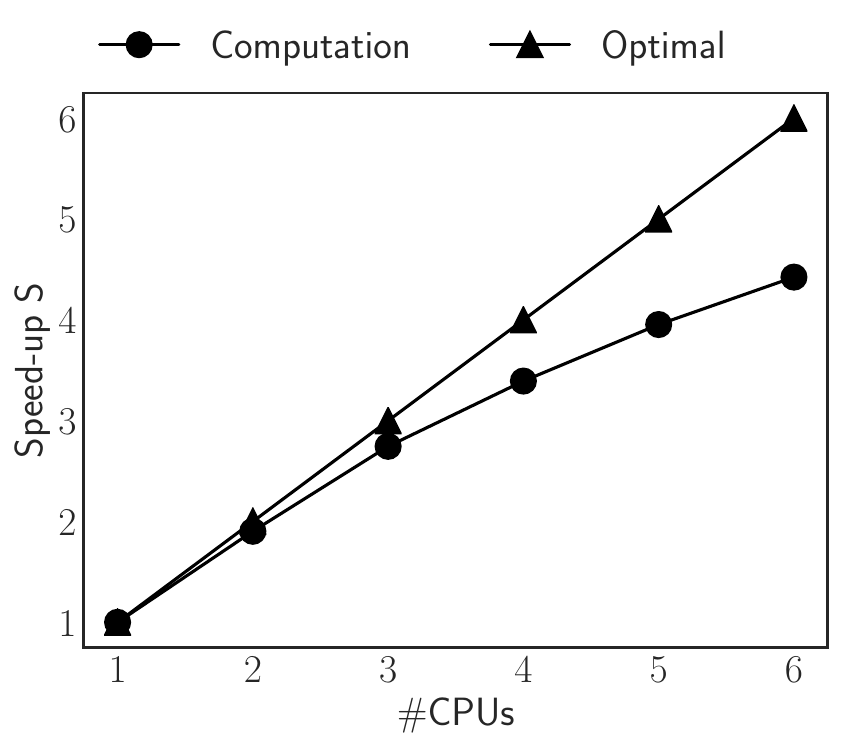}
        \caption{Speed-up $\mathsf{S}(p)=\sfrac{T_1}{T_p}$ with respect to the computational time on one node $T_1$ for $p=[1,2,3,\ldots,6]$ where $p$ is the number of CPUs. Note that strong scaling was used.}
        \label{fig:benchmark:im2:sppedup}
\end{subfigure}
    
\begin{subfigure}[t]{0.6\textwidth}
\includegraphics[width=\textwidth]{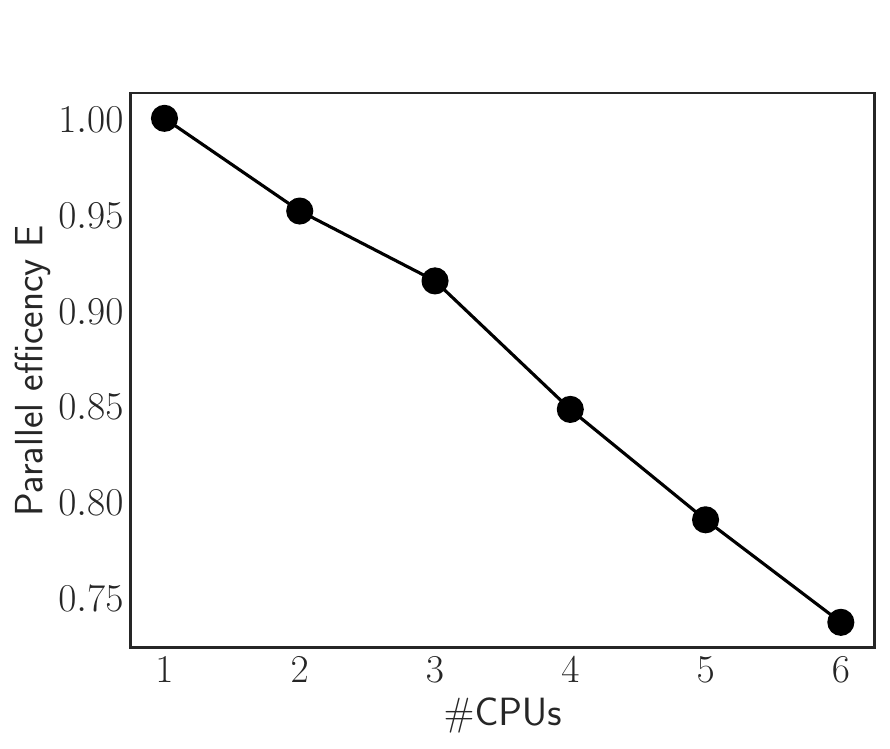}
\caption{Parallel efficiency $\mathsf{E}(p)=\sfrac{S(p)}{p}$ for $p=[1,2,3,\ldots,6]$ where $p$ is the number of CPUs. The parallel efficiency is some indication for the fraction of time for which the CPU is doing some computation.}
\label{fig:benchmark:im2:efficency}
\end{subfigure}
\caption{The speed-up $\mathsf{S}$ (\subref{fig:benchmark:im2:sppedup}) and parallel efficiency $\mathsf{E}$ (\subref{fig:benchmark:im2:efficency}) for the test case in Section~\ref{sec:validation:im:2} using the implicit scheme. Note that all presented results are for strong scaling.}
\label{fig:benchmark:im2}
\end{figure}    
    
\noindent
The speed-up $\mathsf{S}(p)=\sfrac{T_1}{T_p}$~\cite{rodgers1985improvements} with respect to the computational time on one node $T_1$ is shown in Figure~\ref{fig:benchmark:im2:sppedup} for $p=[1,2,3,\ldots,6]$ where $p$ is the number of CPUs. The straight line shows the optimal speed-up, meaning that we assume if we go from one CPU to two CPUs the code gets two time faster and so. The lines with the circle markers shows the speed-up with respect to one single CPU. Here, up to three CPUs, we are close to the optimal speed-up. Later we divergence from the optimal speed-up probably due to the strong scaling and the fixed problem size. In addition, the parallel efficiency $\mathsf{E}(p)=\sfrac{S(p)}{p}$~\cite{rodgers1985improvements} is shown in Figure~\ref{fig:benchmark:im2:efficency} for $p=[1,2,3,\ldots,6]$ where $p$ is the number of CPUs. The parallel efficiency is some indication for the fraction of time for which the CPU is doing some computation. The parallel efficiency is above $0.9$ for up to three CPUs. For four to five CPUs the parallel efficiency decays to $0.8$ and for six CPUs the efficiency is around $0.75$. \yellow{More} sophisticated execution policies (\emph{e.g.}\ \yellow{dynamic or adaptive chunk size}) could be applied, to decrease the computational time. Note that only parallel for loops, synchronization, and futurization were utilized for parallelization.


\subsection{Explicit}
The setup presented in Section~\ref{sec:validation:exp:2D} was discretized with $196249$ nodes and an horizon of $0.05$\,\si{\meter} and $m_d=20$ as chosen. The benchmark was run on CentOS $7$ with kernel $3$.$10$.$0$ on Intel(R) Xeon(R) CPU E$5$-$2690$ @ $2$.$90$GHz. HPX (version $82f7b281$) and PeridynamicHPX were compiled with gcc $7$.$2$, boost $1$.$61$ and blaze $3.2$ libraries.\\

\begin{figure}[tbp]
\centering
\begin{subfigure}[c]{0.6\textwidth}
\includegraphics[width=\textwidth]{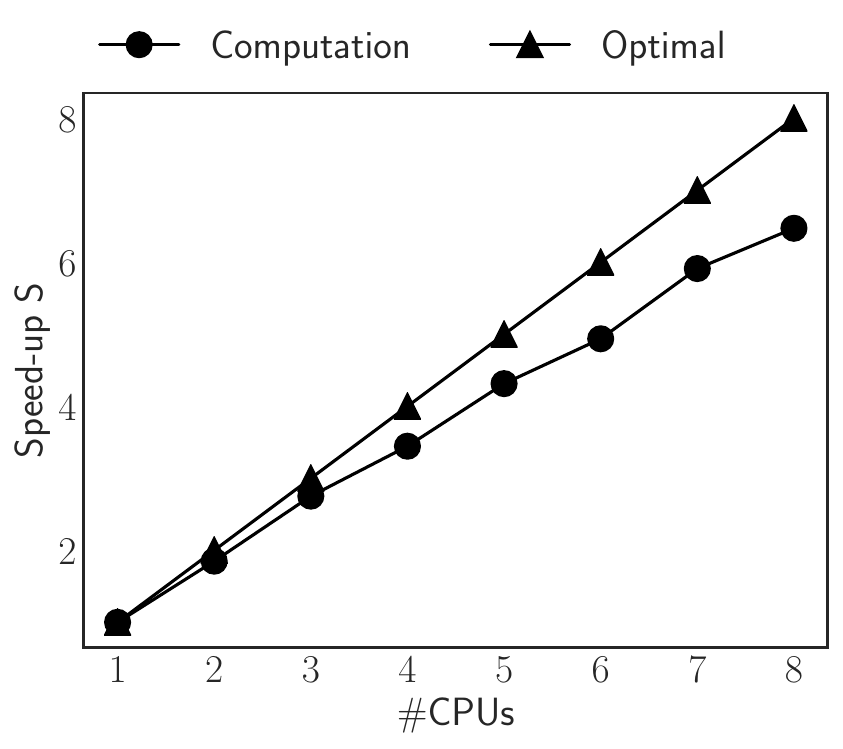}
\caption{Speed-up $\mathsf{S}(\text{CPUs})=\sfrac{T_1}{T_\text{CPUs}}$ with respect to the computational time on one node $T_1$ for $\text{CPUS}=[1,2,3,\ldots,8]$ where $p$ is the number of CPUs.}
\label{fig:benchmark:ex2:sppedup}
\end{subfigure}

\begin{subfigure}[t]{0.6\textwidth}
\includegraphics[width=\textwidth]{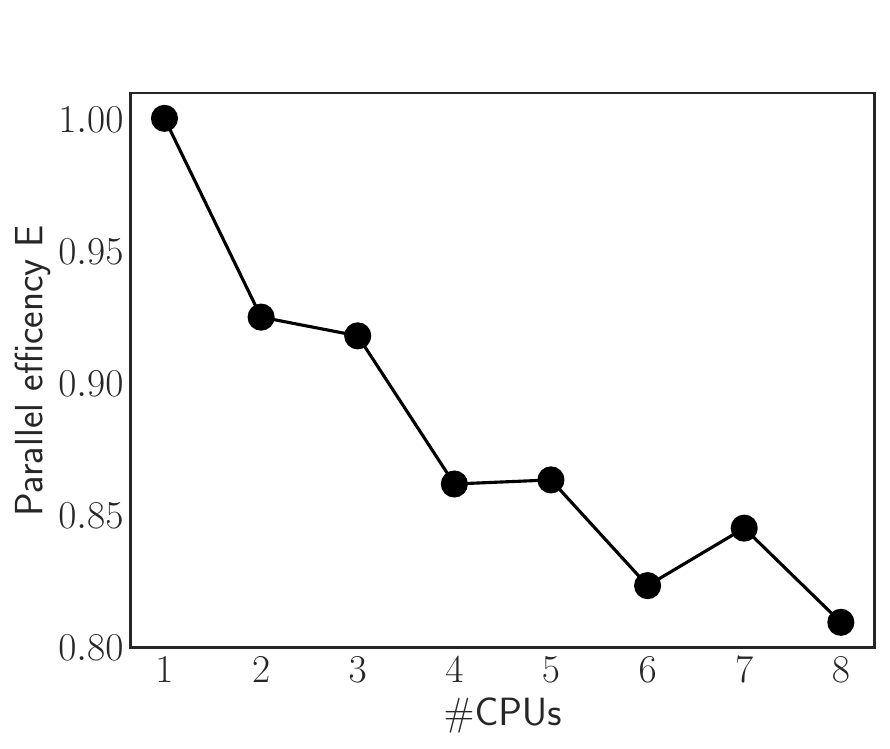}
\caption{Parallel efficiency $\mathsf{E}(p)=\sfrac{S(p)}{p}$ for $p=[1,2,3,\ldots,6]$ where $p$ is the number of CPUs. The parallel efficiency is some indication for the fraction of time for which the CPU is doing some computation.}
\label{fig:benchmark:ex2:efficency}
\end{subfigure}
\caption{The speed-up $S$ (\subref{fig:benchmark:ex2:sppedup}) and parallel efficiency $E$. Note that all presented results are for strong scaling.}
\label{fig:benchmark:ex2}
\end{figure}

\noindent
The speed-up $\mathsf{S}(p)=\sfrac{T_1}{T_p}$~\cite{rodgers1985improvements} with respect to the computational time on one node $T_1$ is shown in Figure~\ref{fig:benchmark:im2:sppedup} for $p=[1,2,3,\ldots,8]$ where $p$ is the number of CPUs. The straight line shows the optimal speed-up, meaning that we assume if we go from one CPU to two CPUs the code gets two time faster and so. The lines with the circle markers shows the speed-up with respect to one single CPU. Here, up to three CPUs, we are close to the optimal speed-up. Later we divergence from the optimal speed-up probably due to the strong scaling and the fixed problem size. In addition, the parallel efficiency $\mathsf{E}(p)=\sfrac{S(p)}{p}$~\cite{rodgers1985improvements} is shown in Figure~\ref{fig:benchmark:ex2:efficency} for $p=[1,2,3,\ldots,8]$ where $p$ is the number of CPUs. The parallel efficiency is some indication for the fraction of time for which the CPU is doing some computation. The parallel efficiency is above $0.9$ for up to three CPUs. For four to five CPUs the parallel efficiency decays to $0.85$ and for six CPUs the efficiency is around $0.82$. For seven CPUs the efficiency increases again to $0.84$. For eight CPUs the efficiency decays again to $0.81$. The behavior of the efficiency rate might be related to the fix problem six of the strong scaling. \yellow{More} sophisticated execution policies (\emph{e.g.}\ dynamic or adaptive chunk size) could be applied and larger problem sizes, to decrease the computational time. Note that only parallel for loops, synchronization, and futurization were utilized for parallelization.

\section{Conclusion}
\label{sec:conclusion}
Bond-based and state-based elastic peridynamic material models and implicit and explicit time integration schemes were implemented within a asynchronous many task run time system. These run time systems, like HPX, are essential for utilizing the full performance of the cluster with many cores on a single node.\\

\noindent
One important part of the design was the modular expandability for the extensions. New material models can be integrated into the code by inheriting the functions of an abstract class. Consequently, only the material specific functionality, like forces or strain, is provided by the user and implementation details for the parallelism and concurrency are hidden from the user as much as possible. Additional HPX-related functionality needs to be considered for the extension to other integration schemes.\\
Materials models and the different time integration schemes were validated against theoretical solutions and classical continuum mechanics. All are in good agreement with reference solutions. The convergence rate was shown to be closer to theoretical value, which suggests that the code behaves as expected. The solutions converge to the exact solution at a rate of 1.
%
\\

\noindent
The code scaling with respect to computational resource is important and our benchmark results show that the scaling achieved is very close to the theoretical estimates. In addition, is the speed-up $\mathsf{S}$ and the parallel efficiency $\mathsf{E}$ are reasonable for a non-optimized code using default execution policies. Both integration schemes were compared against theoretical estimations. The trend of the theoretical estimates fits with measured computational time and both integration schemes scale with increasing amounts of CPUs. These results were obtained by the default execution policies without any optimization.



\clearpage
\newpage
  \bibliographystyle{abbrv} 
  \bibliography{bib.bib}

\begin{thebibliography}{10}

\bibitem{starpu}
C.~Augonnet, S.~Thibault, R.~Namyst, and P.-A. Wacrenier.
\newblock {StarPU: A unified platform for task scheduling on heterogeneous
  multicore architectures}.
\newblock In {\em {European Conference on Parallel Processing}}, pages
  863--874. Springer, 2009.

\bibitem{ac:2017}
J.~Biddiscombe, T.~Heller, A.~Bikineev, and H.~Kaiser.
\newblock {Zero Copy Serialization using RMA in the Distributed Task-Based HPX
  runtime}.
\newblock In {\em {14th International Conference on Applied Computing}}. IADIS,
  International Association for Development of the Information Society, 2017.

\bibitem{intelcilkplus}
R.~D. Blumofe, C.~F. Joerg, B.~C. Kuszmaul, C.~E. Leiserson, K.~H. Randall, and
  Y.~Zhou.
\newblock {Cilk}: An efficient multithreaded runtime system.
\newblock In {\em Proceedings of the Fifth ACM SIGPLAN Symposium on Principles
  and Practice of Parallel Programming (PPoPP)}, pages 207--216, Santa Barbara,
  California, July 1995.

\bibitem{chamberlain07parallelprogrammability}
B.~L. Chamberlain, D.~Callahan, and H.~P. Zima.
\newblock Parallel programmability and the {Chapel} language.
\newblock {\em International Journal of High Performance Computing
  Applications}, 21:291--312, 2007.

\bibitem{Charles:2005:XOA:1094811.1094852}
P.~Charles, C.~Grothoff, V.~Saraswat, C.~Donawa, A.~Kielstra, K.~Ebcioglu,
  C.~von Praun, and V.~Sarkar.
\newblock {X10: an object-oriented approach to non-uniform cluster computing}.
\newblock In {\em Proceedings of the 20th annual ACM SIGPLAN conference on
  Object-oriented programming, systems, languages, and applications}, OOPSLA
  '05, pages 519--538, New York, NY, USA, 2005. ACM.

\bibitem{chen2011continuous}
X.~Chen and M.~Gunzburger.
\newblock Continuous and discontinuous finite element methods for a
  peridynamics model of mechanics.
\newblock {\em Computer Methods in Applied Mechanics and Engineering},
  200(9):1237--1250, 2011.

\bibitem{dai2019piz}
G.~Dai{\ss}, P.~Amini, J.~Biddiscombe, P.~Diehl, J.~Frank, K.~Huck, H.~Kaiser,
  D.~Marcello, D.~Pfander, and D.~Pf{\"u}ger.
\newblock {From Piz Daint to the stars: simulation of stellar mergers using
  high-level abstractions}.
\newblock In {\em {Proceedings of the International Conference for High
  Performance Computing, Networking, Storage and Analysis}}, pages 1--37, 2019.

\bibitem{d2019review}
M.~D'Elia, X.~Li, P.~Seleson, X.~Tian, and Y.~Yu.
\newblock A review of local-to-nonlocal coupling methods in nonlocal diffusion
  and nonlocal mechanics.
\newblock {\em arXiv preprint arXiv:1912.06668}, 2019.

\bibitem{Diehl:2012}
P.~Diehl.
\newblock {I}mplementierung eines {P}eridynamik-{V}erfahrens auf {GPU}.
\newblock Diplomarbeit, {I}nstitute of {P}arallel and {D}istributed {S}ystems,
  {U}niversity of {S}tuttgart, 2012.

\bibitem{diehl_2020_2d}
P.~Diehl.
\newblock Validation 2d, Jul 2020.

\bibitem{Diehl2020_validation_1d}
P.~Diehl.
\newblock {Validation of a one-dimensional bar}.
\newblock 5 2020.

\bibitem{Diehl2019review}
P.~Diehl, S.~Prudhomme, and M.~L{\'e}vesque.
\newblock {A Review of Benchmark Experiments for the Validation of Peridynamics
  Models}.
\newblock {\em {Journal of Peridynamics and Nonlocal Modeling}}, Feb 2019.

\bibitem{CarterEdwards20143202}
H.~C. Edwards, C.~R. Trott, and D.~Sunderland.
\newblock {Kokkos: Enabling manycore performance portability through
  polymorphic memory access patterns}.
\newblock {\em {Journal of Parallel and Distributed Computing}}, 74(12):3202 --
  3216, 2014.

\bibitem{emmrich2007peridynamic}
E.~Emmrich and O.~Weckner.
\newblock The peridynamic equation and its spatial discretisation.
\newblock {\em Mathematical Modelling and Analysis}, 12(1):17--27, 2007.

\bibitem{grun2015brief}
P.~Grun, S.~Hefty, S.~Sur, D.~Goodell, R.~D. Russell, H.~Pritchard, and J.~M.
  Squyres.
\newblock {A brief introduction to the openfabrics interfaces-a new network api
  for maximizing high performance application efficiency}.
\newblock In {\em {2015 IEEE 23rd Annual Symposium on High-Performance
  Interconnects}}, pages 34--39. {IEEE}, 2015.

\bibitem{heller2013using}
T.~Heller, H.~Kaiser, A.~Sch{\"a}fer, and D.~Fey.
\newblock {Using HPX and LibGeoDecomp for scaling HPC applications on
  heterogeneous supercomputers}.
\newblock In {\em {Proceedings of the Workshop on Latest Advances in Scalable
  Algorithms for Large-Scale Systems}}, page~1. ACM, 2013.

\bibitem{operationbell17}
T.~Heller, B.~A. Lelbach, K.~A. Huck, J.~Biddiscombe, P.~Grubel, A.~E. Koniges,
  M.~Kretz, D.~Marcello, D.~Pfander, A.~Serio, J.~Frank, G.~C. Clayton,
  D.~Pflüger, D.~Eder, and H.~Kaiser.
\newblock {Harnessing billions of tasks for a scalable portable hydrodynamic
  simulation of the merger of two stars}.
\newblock {\em The International Journal of High Performance Computing
  Applications}, 0(0), 0.

\bibitem{iglberger2012expression}
K.~Iglberger, G.~Hager, J.~Treibig, and U.~R{\"u}de.
\newblock Expression templates revisited: a performance analysis of current
  methodologies.
\newblock {\em SIAM Journal on Scientific Computing}, 34(2):C42--C69, 2012.

\bibitem{6266939}
K.~Iglberger, G.~Hager, J.~Treibig, and U.~R{\"u}de.
\newblock High performance smart expression template math libraries.
\newblock In {\em 2012 International Conference on High Performance Computing
  Simulation (HPCS)}, pages 367--373, July 2012.

\bibitem{javili2019peridynamics}
A.~Javili, R.~Morasata, E.~Oterkus, and S.~Oterkus.
\newblock Peridynamics review.
\newblock {\em {Mathematics and Mechanics of Solids}}, 24(11):3714--3739, 2019.

\bibitem{CMPer-JhaLipton}
P.~K. Jha and R.~Lipton.
\newblock Numerical analysis of nonlocal fracture models in h\"older space.
\newblock {\em SIAM Journal on Numerical Analysis}, 56(2):906--941, 2018.

\bibitem{CMPer-JhaLipton2}
P.~K. Jha and R.~Lipton.
\newblock Numerical convergence of nonlinear nonlocal continuum models to local
  elastodynamics.
\newblock {\em International Journal for Numerical Methods in Engineering},
  114(13):1389--1410, 2018.

\bibitem{CMPer-JhaLipton5}
P.~K. Jha and R.~Lipton.
\newblock Numerical convergence of finite difference approximations for state
  based peridynamic fracture models.
\newblock {\em Computer Methods in Applied Mechanics and Engineering (2019)},
  March 2019.

\bibitem{kaiser2009parallex}
H.~Kaiser, M.~Brodowicz, and T.~Sterling.
\newblock {Parallex an advanced parallel execution model for scaling-impaired
  applications}.
\newblock In {\em {Parallel Processing Workshops, 2009. ICPPW'09.}}, pages
  394--401. IEEE, 2009.

\bibitem{Kaiser2020}
H.~Kaiser, P.~Diehl, A.~S. Lemoine, B.~A. Lelbach, P.~Amini, A.~Berge,
  J.~Biddiscombe, S.~R. Brandt, N.~Gupta, T.~Heller, K.~Huck, Z.~Khatami,
  A.~Kheirkhahan, A.~Reverdell, S.~Shirzad, M.~Simberg, B.~Wagle, W.~Wei, and
  T.~Zhang.
\newblock {HPX - The C++ Standard Library for Parallelism and Concurrency}.
\newblock {\em {Journal of Open Source Software}}, 5(53):2352, 2020.

\bibitem{kaiser2014hpx}
H.~Kaiser, T.~Heller, B.~Adelstein-Lelbach, A.~Serio, and D.~Fey.
\newblock Hpx: A task based programming model in a global address space.
\newblock In {\em {Proceedings of the 8th International Conference on
  Partitioned Global Address Space Programming Models}}, page~6. ACM, 2014.

\bibitem{Kaiser:2015:HPL:2832241.2832244}
H.~Kaiser, T.~Heller, D.~Bourgeois, and D.~Fey.
\newblock Higher-level parallelization for local and distributed asynchronous
  task-based programming.
\newblock In {\em Proceedings of the First International Workshop on Extreme
  Scale Programming Models and Middleware}, ESPM '15, pages 29--37, New York,
  NY, USA, 2015. ACM.

\bibitem{khatami2016massively}
Z.~Khatami, H.~Kaiser, P.~Grubel, A.~Serio, and J.~Ramanujam.
\newblock A massively parallel distributed n-body application implemented with
  hpx.
\newblock In {\em {2016 7th Workshop on Latest Advances in Scalable Algorithms
  for Large-Scale Systems (ScalA)}}, pages 57--64. IEEE, 2016.

\bibitem{kilic_peridynamic_nodate}
B.~Kilic.
\newblock {\em Peridynamic {Theory} for {Progressive} {Failure} {Prediction} in
  {Homogeneous} and {Heterogeneous} {Materials}}.
\newblock The University of Arizona., 2008.

\bibitem{Kunin2011}
I.~A. Kunin.
\newblock {\em Elastic Media with Microstructure I: One-Dimensional Models
  (Springer Series in Solid-State Sciences)}.
\newblock Springer, softcover reprint of the original 1st ed. 1982 edition, 12
  2011.

\bibitem{Kunin2012}
I.~A. Kunin.
\newblock {\em Elastic Media with Microstructure II: Three-Dimensional Models
  (Springer Series in Solid-State Sciences)}.
\newblock Springer, softcover reprint of the original 1st ed. 1983 edition, 1
  2012.

\bibitem{CMPer-Lipton3}
R.~Lipton.
\newblock Dynamic brittle fracture as a small horizon limit of peridynamics.
\newblock {\em Journal of Elasticity}, 117(1):21--50, 2014.

\bibitem{CMPer-Lipton}
R.~Lipton.
\newblock Cohesive dynamics and brittle fracture.
\newblock {\em Journal of Elasticity}, 124(2):143--191, 2016.

\bibitem{CMPer-Lipton2}
R.~P. Lipton, R.~B. Lehoucq, and P.~K. Jha.
\newblock Complex fracture nucleation and evolution with nonlocal
  elastodynamics.
\newblock {\em Journal of Peridynamics and Nonlocal Modeling}, 1(2):122--130,
  Oct 2019.

\bibitem{Littlewood2015}
D.~J. Littlewood.
\newblock {Roadmap for Peridynamic Software Implementation}.
\newblock Technical Report 2015-9013, {Sandia National Laboratories}, 2015.

\bibitem{MOSSAIBY20171856}
F.~Mossaiby, A.~Shojaei, M.~Zaccariotto, and U.~Galvanetto.
\newblock Opencl implementation of a high performance 3d peridynamic model on
  graphics accelerators.
\newblock {\em Computers \& Mathematics with Applications}, 74(8):1856 -- 1870,
  2017.

\bibitem{openmpspec}
{OpenMP Architecture Review Board}.
\newblock {OpenMP V5.0 Specification}, 2018.
\newblock
  \url{https://www.openmp.org/wp-content/uploads/OpenMPRef-5.0-0519-web.pdf}.

\bibitem{Parks2012}
M.~Parks, D.~Littlewood, J.~Mitchell, and S.~Silling.
\newblock {Peridigm Users’ Guide}.
\newblock Technical Report SAND2012-7800, {Sandia National Laboratories}, 2012.

\bibitem{CMPer-Parks}
M.~L. Parks, R.~B. Lehoucq, S.~J. Plimpton, and S.~A. Silling.
\newblock Implementing peridynamics within a molecular dynamics code.
\newblock {\em Computer Physics Communications}, 179(11):777--783, 2008.

\bibitem{Pfander:2018:AOS:3204919.3204938}
D.~Pfander, G.~Dai\ss, D.~Marcello, H.~Kaiser, and D.~Pfl\"{u}ger.
\newblock {Accelerating Octo-Tiger: Stellar Mergers on Intel Knights Landing
  with HPX}.
\newblock In {\em {Proceedings of the International Workshop on OpenCL}}, IWOCL
  '18, pages 19:1--19:8, New York, NY, USA, 2018. ACM.

\bibitem{intelispc}
M.~{Pharr} and W.~R. {Mark}.
\newblock ispc: A spmd compiler for high-performance cpu programming.
\newblock In {\em {2012 Innovative Parallel Computing (InPar)}}, pages 1--13,
  2012.

\bibitem{inteltbb}
C.~Pheatt.
\newblock Intel® threading building blocks.
\newblock {\em J. Comput. Sci. Coll.}, 23(4):298, Apr. 2008.

\bibitem{4777912}
I.~Raicu, I.~T. Foster, and Y.~Zhao.
\newblock Many-task computing for grids and supercomputers.
\newblock In {\em 2008 Workshop on Many-Task Computing on Grids and
  Supercomputers}, pages 1--11, Nov 2008.

\bibitem{rodgers1985improvements}
D.~P. Rodgers.
\newblock Improvements in multiprocessor system design.
\newblock {\em {ACM SIGARCH Computer Architecture News}}, 13(3):225--231, 1985.

\bibitem{ross2008cpu}
P.~E. Ross.
\newblock Why cpu frequency stalled.
\newblock {\em IEEE Spectrum}, 45(4), 2008.

\bibitem{sadd2009elasticity}
M.~H. Sadd.
\newblock {\em Elasticity: theory, applications, and numerics}.
\newblock Academic Press, 2009.

\bibitem{seleson2016convergence}
P.~Seleson and D.~J. Littlewood.
\newblock Convergence studies in meshfree peridynamic simulations.
\newblock {\em Computers \& Mathematics with Applications}, 71(11):2432--2448,
  2016.

\bibitem{Seo2018ArgobotsAL}
S.~Seo, A.~Amer, P.~Balaji, C.~Bordage, G.~Bosilca, A.~Brooks, P.~H. Carns,
  A.~Castellx00F3, D.~Genet, T.~H{\'e}rault, S.~Iwasaki, P.~Jindal, L.~V.
  Kalx00E9, S.~Krishnamoorthy, J.~Lifflander, H.~Lu, E.~Meneses, M.~Snir,
  Y.~Sun, K.~Taura, and P.~H. Beckman.
\newblock Argobots: A lightweight low-level threading and tasking framework.
\newblock {\em IEEE Transactions on Parallel and Distributed Systems},
  29:512--526, 2018.

\bibitem{CMPer-Silling}
S.~A. Silling.
\newblock Reformulation of elasticity theory for discontinuities and long-range
  forces.
\newblock {\em Journal of the Mechanics and Physics of Solids}, 48(1):175--209,
  2000.

\bibitem{silling2005meshfree}
S.~A. Silling and E.~Askari.
\newblock A meshfree method based on the peridynamic model of solid mechanics.
\newblock {\em Computers \& structures}, 83(17):1526--1535, 2005.

\bibitem{silling2007peridynamic}
S.~A. Silling, M.~Epton, O.~Weckner, J.~Xu, and E.~Askari.
\newblock Peridynamic states and constitutive modeling.
\newblock {\em Journal of Elasticity}, 88(2):151--184, 2007.

\bibitem{strang1980}
G.~Strang.
\newblock {\em Linear Algebra and Its Applications)}.
\newblock Academic Press, Inc., 1980.

\bibitem{sutter2005free}
H.~Sutter.
\newblock The free lunch is over: A fundamental turn toward concurrency in
  software.
\newblock {\em Dr. Dobb’s journal}, 30(3):202--210, 2005.

\bibitem{cxx11_standard}
{The C++ Standards Committee}.
\newblock {ISO International Standard ISO/IEC 14882:2011, Programming Language
  C++}.
\newblock Technical report, {Geneva, Switzerland: International Organization
  for Standardization (ISO).}, 2011.
\newblock \url{http://www.open-std.org/jtc1/sc22/wg21}.

\bibitem{cxx17_standard}
{The C++ Standards Committee}.
\newblock {ISO International Standard ISO/IEC 14882:2017, Programming Language
  C++}.
\newblock Technical report, {Geneva, Switzerland: International Organization
  for Standardization (ISO)}, 2017.
\newblock \url{http://www.open-std.org/jtc1/sc22/wg21}.

\bibitem{thoman2018taxonomy}
P.~Thoman, K.~Dichev, T.~Heller, R.~Iakymchuk, X.~Aguilar, K.~Hasanov,
  P.~Gschwandtner, P.~Lemarinier, S.~Markidis, H.~Jordan, et~al.
\newblock A taxonomy of task-based parallel programming technologies for
  high-performance computing.
\newblock {\em {The Journal of Supercomputing}}, 74(4):1422--1434, 2018.

\bibitem{WANG20127730}
H.~Wang and H.~Tian.
\newblock {A fast Galerkin method with efficient matrix assembly and storage
  for a peridynamic model}.
\newblock {\em {Journal of Computational Physics}}, 231(23):7730 -- 7738, 2012.

\bibitem{weckner2005numerical}
O.~Weckner and E.~Emmrich.
\newblock Numerical simulation of the dynamics of a nonlocal, inhomogeneous,
  infinite bar.
\newblock {\em Journal of Computational and Applied Mechanics}, 6(2):311--319,
  2005.

\bibitem{qthreads}
K.~B. Wheeler, R.~C. Murphy, and D.~Thain.
\newblock Qthreads: An api for programming with millions of lightweight
  threads.
\newblock In {\em 2008 IEEE International Symposium on Parallel and Distributed
  Processing}, pages 1--8. IEEE, 2008.

\end{thebibliography}


\end{document}